\documentstyle[12pt]{article}

\begin{document}

\def\thefootnote{\fnsymbol{footnote}}

{\it University of Shizuoka}

\hspace*{10cm} {\bf US-95-04}\\[-.3in]

\hspace*{10cm} {\bf EHU-95-03}\\[-.3in]

\hspace*{10cm} {\bf Revised Version}\\[-.3in]

\hspace*{10cm} {\bf  January 1996}\\[.2in]

\begin{center}

{\large\bf U(3)-Family Nonet Higgs Boson}\\[.1in]
{\large\bf and its Phenomenology}\footnote{
hep-ph/9505333: To be published in Z.~Phys.~C.}\\[.2in]

{\bf Yoshio Koide\footnote{
E-mail address: koide@u-shizuoka-ken.ac.jp}} \\[.1in]

Department of Physics, University of Shizuoka \\[.1in] 
52-1 Yada, Shizuoka 422, Japan \\[.1in]

and \\[.1in]

{\bf Morimitu Tanimoto\footnote{
E-mail address: tanimoto@edserv.ed.ehime-u.ac.jp}}
\\[.1in]

Science Education Laboratory, Ehime University, 
Matusyama 790, Japan\\[.5in]

{\large\bf Abstract}\\[.1in]
\end{center}

\begin{quotation}
In a model where quark and lepton masses and family-mixings are 
caused not by a variety of  Yukawa couplings $y_{ij}$
 ($i,j=1,2,3$: family indices) with one vacuum expectation value (VEV) 
$v=\langle\phi_L^0\rangle_0$, but by a variety of VEV's of 
a U(3)-family nonet Higgs boson $\phi_L$, 
$v_{i}^j=\langle\phi_{L i}^{0j}\rangle_0$, with a  single coupling constant,
the following problems are investigated:
what constraints on the Higgs potential are imposed in order 
to provide realistic quark and lepton mass spectra and mixings 
and what constraints on the Higgs boson masses are required 
in order to suppress unwelcome flavor-changing neutral current
effects.
Lower bounds of the physical Higgs boson masses  of $\phi_L$ 
are deduced from the present 
experimental data and new physics from the present scenario 
is speculated.
\end{quotation}

\newpage

\noindent{\bf 1. Introduction}

\vglue.1in

One of our dissatisfactions with the standard model is that for the 
explanation of the mass spectra of quarks and leptons, 
we are obliged to choose the coefficients $y_{ij}^f$ 
in the Yukawa coupling
$\sum_f \sum_{i,j} \overline{f}_L^i f_{jR} \langle\phi^0\rangle_0$ 
($f=\nu,e,u,d$, and $i,j$ are family indices) ``by hand". 
If we could understand the mass spectra from the vacuum expectation 
values (VEV's)  $\langle\phi_i^{0j}\rangle_0$ of 
U(3)$_{family}$ [1] nonet Higgs fields 
which couple with fermions as 
$\sum_f\sum_{i,j}\overline{f}_L^i\langle\phi_i^{0j}\rangle_0 f_{jR}$, 
we would be happy.
Unfortunately, however, we know that the mass spectra of up- and 
down-quarks and charged leptons are not identical 
and the Kobayashi-Maskawa [2] (KM) matrix is not a unit matrix.
Moreover, we know that in such multi-Higgs models, in general, 
flavor changing neutral currents (FCNC) appear unfavorably.

In the present paper, on the basis of a model where quark and 
lepton masses and family-mixings are 
caused not by a variety of  Yukawa couplings $y_{ij}$
 ($i,j=1,2,3$: family indices) with one vacuum expectation value (VEV) 
$v=\langle\phi_L^0\rangle_0$, but by a variety of VEV's $v_{i}^j=
\langle\phi_{L i}^{0j}\rangle_0$ with a  single coupling constant,
we investigate the following problems:
what constraints on the Higgs potential are imposed in order 
to provide realistic quark and lepton mass spectra and KM mixings 
and what constraints on the Higgs boson masses are required 
in order to suppress unwelcome FCNC  effects.
It will be concluded that a special form of the Higgs potential 
$V(\phi_L)$, which leads to realistic quark and lepton mass 
spectra, can safely suppress unwelcome FCNC and the present 
experimental data put lower bounds of a few TeV on the 
physical Higgs boson masses.

The model we discuss is  a seesaw-type quark and lepton mass matrix 
model [3], where the $6\times 6$ mass matrix for fermions $f$ and 
$F$ are given by 
$$
(\overline{f}\ \overline{F})_L
\left(
\begin{array}{ccc}
0 & m_L \\
m_R & M_F
\end{array} \right) \left(
\begin{array}{c}
f \\
F
\end{array} \right)_R \ , \eqno(1.1)
$$
where $f_L=({\bf 2},{\bf 1})$, $f_R=({\bf 1},{\bf 2})$, 
$F_L=({\bf 1},{\bf 1})$ and $F_R=({\bf 1},{\bf 1})$ of 
SU(2)$_L \times$SU(2)$_R$. 
We assume that $3\times 3$ matrices $m_L$ and $m_R$ are universal 
for $f=u$, $d$, $\nu$ and $e$ (up-quark-, down-quark-, neutrino- 
and charged lepton-sectors),
so that differences between quark and lepton sectors and between 
up- and down-sectors come only from the differences of $M_F$.
We assume that the structure of $M_F$ is simply given by 
[(unit matrix)+ $b_f$ (a rank-one matrix)], 
where $b_f$ is a complex parameter depending on $f$ 
(up- or down- and quark or lepton sectors). 
The SU(2)$_L$ [SU(2)$_R$] symmetry breaking matrix $m_L$ [$m_R$] 
is given by $y_L\langle\phi_L^0\rangle_0$ 
[$y_R\langle\phi_R^0\rangle_0$], where $\phi_L$ [$\phi_R$] belongs to 
({\bf 2},{\bf 1},{\bf 8}+{\bf 1}) [({\bf1},{\bf 2},{\bf 8+1})] 
of SU(2)$_L \times$SU(2)$_R\times$U(3)$_{family}$. 
Note that the U(3)-family symmetry is badly broken by the heavy 
fermion mass matrix $M_F$, as we state in Sect.2.

Generally, the diagonalization of the mass matrix (1.1) transforms 
the vertex 
$$
(\overline{f}\ \overline{F})_L
\left(
\begin{array}{ccc}
0 & \Gamma_{12} \\
\Gamma_{21} & \Gamma_{22}
\end{array} \right) \left(
\begin{array}{c}
f \\
F
\end{array} \right)_R \ , \eqno(1.2)
$$
into 
$$
(\overline{f}'\ \overline{F}')_L
\left(
\begin{array}{ccc}
\Gamma'_{11} & \Gamma'_{12} \\
\Gamma'_{21} & \Gamma'_{22}
\end{array} \right) \left(
\begin{array}{c}
f' \\
F'
\end{array} \right)_R \ , \eqno(1.3)
$$
where  $\Gamma_{12}=y_L\phi^0_L$ and so on, and 
$(f',F')$ are mass eigenstates, so that the vertex 
$\Gamma'_{11}$ is not $\Gamma'_{11}=0$ any longer.
(The details are discussed in Sect.3.)
Since the physical Higgs bosons $\phi_L$ are sufficiently light 
compared with the other Higgs bosons $\phi_R$ and so on, 
the contributions to FCNC in quarks and leptons will be dominated
only by $\phi_L$.
Therefore, in the present paper, we will concentrate 
our study on  the Higgs boson $\phi_L$.

In the present paper, as a model of the Higgs boson $\phi_L$, 
we adopt a U(3)-family nonet Higgs boson model [4], which 
was proposed by one of the authors (Y.K.) in order to explain 
a charged lepton mass relation [5]
$$
m_e+m_\mu+m_\tau=\frac{2}{3}(\sqrt{m_e}+\sqrt{m_\mu}+\sqrt{m_\tau})^2
\ , \eqno(1.4)
$$
which  predicts $  m_\tau = 1776.969\pm 0.001$ MeV
for the input values [6] of $m_e$ and $m_\mu$.
He assumed  a U(3)$_{family}$ nonet Higgs boson $\phi$ 
whose potential is given by  
$$V(\phi)=\mu^2{\rm Tr}(\phi\phi^\dagger)
+\frac{1}{2}\lambda\left[{\rm Tr}(\phi\phi^\dagger)\right]^2
+\eta\phi_s\phi_s^*{\rm Tr}(\phi_{oct}\phi_{oct}^\dagger) \ . \eqno(1.5)
$$
Here, for simplicity, the SU(2)$_L$ structure of $\phi$ has been neglected,
and we have expressed the nonet Higgs bosons $\phi_i^j$ by 
the form of $3\times 3$ matrix, 
$$
 \phi=\phi_{oct}+\frac{1}{\sqrt{3}}\phi_s\; {\bf 1}\ ,\eqno(1.6)
$$
where $\phi_{oct}$ is the octet part of $\phi$, i.e., Tr$(\phi_{oct})=0$, 
and {\bf 1} is a $3\times 3$ unit matrix. 
For $\mu^2<0$, conditions for minimizing the potential (1.5) lead to 
the relation
$$
v_s^*v_s = {\rm Tr}\left( v_{oct}^\dagger v_{oct} \right)\ ,  \eqno(1.7) 
$$
together with $v=v^\dagger$, where $v=\langle \phi\rangle_0$, 
$v_{oct}=\langle \phi_{oct}\rangle_0$ and 
$v_s=\langle \phi_s\rangle_0$,
so that we obtain the relation 
$$
{\rm Tr}\left(v^2\right) = 
\frac{2}{3} \left[{\rm Tr}(v)\right]^2 \ . \eqno(1.8)
$$
If we assume a seesaw-like mechanism for charged lepton mass matrix 
$M_e$, $ M_e \simeq m M_E^{-1} m $, with $m\propto v$ and heavy lepton 
mass matrix $M_E\propto {\bf 1}$, we can obtain the mass relation (1.4).

However, the model (1.5) is only a toy model, and here the SU(2)$_L$ 
structure of $\phi$ was not discussed explicitly. 
Moreover, the Higgs potential (1.5) brings unwelcome massless 
physical Higgs bosons into the theory.
In the present paper, we investigate what potential form of $\phi_L$ 
is favorable in order to provide realistic fermion mass spectrum 
without contradicting the present experimental data.
The outline of the model is presented in the next section 2.

In the  section 3, we discuss the Higgs potential $V(\phi_L)$ of 
the U(3)-family nonet Higgs bosons $\phi_L$ ($\phi_R$) under an ansatz
and the conditions for minimizing $V(\phi_L)$.
In Sect.4, we calculate masses of the Higgs boson $\phi_L$, and 
in Sect.5, we estimate a lower bound of the mass of the Higgs bosons 
$\phi_{Li}^j$ ($i\neq j$) from the experimental data of the rare kaon 
decay $K_L\rightarrow e^\pm \mu^\mp$.
Besides, the present model, in general, 
induces FCNC.
In Sect.6, we will estimate  lower bounds of the physical Higgs boson masses 
from the present experimental data of $K^0\overline{K}^0$- and 
$D^0\overline{D}^0$-mixings, and so on.
Finally, in Sect.7, we will speculate a possible new physics 
which is expected from
the present model.

\vglue.3in

\noindent{\bf 2. Outline of the model}

\vglue.1in

In our scenario, we prepare the following fermions:
$f=\ell, q$ ($\ell=(\nu,e)$, $q=(u,d)$) and
$F=N, E, U, D$, which belong to $f_L=({\bf 2},{\bf 1},{\bf 3})$, 
$f_R=({\bf 1},{\bf 2},{\bf 3})$, $F_L=({\bf 1},{\bf 1},{\bf 3})$, and 
$F_R=({\bf 1},{\bf 1},{\bf 3})$ of 
SU(2)$_L\times$SU(2)$_R\times$U(3)$_{family}$, respectively. 
Here, SU(2)$_L\times$SU(2)$_R$ are gauged, but 
U(3)$_{family}$ is not gauged.
The global symmetry U(3)$_{family}$ will  be broken 
not spontaneously, but explicitly.
Up- and down-heavy fermions, $F^{up}$ and $F^{down}$, are distinguished 
by hypercharge  $Y$  (note that $Y\neq B-L$ for the heavy fermions): 
Hypercharges of the heavy fermions $(N,E)$ and $(U,D)$ take the values 
$(0,-2)$ and $(4/3,-2/3)$, respectively. 
The quantum numbers of those fields are listed in Table I.

In the present model, differently from the standard 
SU(2)$_L\times$SU(2)$_R\times$U(1)$_{B-L}$ model, 
we do not consider Higgs scalar fields which 
belong to $({\bf 2}, {\bf 2})$ of SU(2)$_L\times$SU(2)$_R$, 
so that there are no Higgs fields which couple with 
$\overline{f}f$ at tree level. 
We assume only the following Yukawa interactions:
$$
H_{Yukawa}=y_0 \sum_{i,j}\overline{F}^i\left[\delta_i^j\Phi_0
+3b_f( \Phi_X)_i^j \right] F_j 
$$
$$
+y_L \sum_{i,j}\left[\overline{f}_L^i(\phi_L)_i^j F_{Rj}^{down} 
+\overline{f}_L^i(\widetilde{\phi}_L)_i^j F_{Rj}^{up} + {\rm h.c.}\right] 
+(L\leftrightarrow R) \ , \eqno(2.1) 
$$
where $\phi=(\phi^+,\phi^0)$ and 
$\widetilde{\phi}=(\overline{\phi}^0,-\phi^-)$.
The scalar fields $\phi_L$ and $\phi_R$ belong to 
$({\bf 2}, {\bf 1},{\bf 8}+{\bf 1})$ and $({\bf 1}, {\bf 2},{\bf 8}+{\bf 1})$ 
of SU(2)$_L\times$SU(2)$_R\times$U(3)$_{family}$, respectively, 
and the VEV's $\langle \phi^0_L\rangle_0$ and 
$\langle \phi^0_R\rangle_0$ provide left- and right-handed weak boson masses 
$m(W_L)$ and $m(W_R)$, respectively.
The fields $\Phi_0$ and $\Phi_X$ which belong to ({\bf 1},{\bf 1},{\bf 1})
and ({\bf 1},{\bf 1},{\bf 8+1}), respectively, do not 
contribute to weak boson masses $m(W_L)$ and $m(W_R)$, but play only 
a role of providing extremely large masses for vector-like fermions $F$. 
Then, the mass matrices for fermions $(f, F)$ are given by 
$$
(\overline{f}\ \overline{F})_L^{up}
\left(
\begin{array}{ccc}
0 & m_L^\dagger \\
m_R & M_F
\end{array} \right) \left(
\begin{array}{c}
f \\
F
\end{array} \right)_R^{up}
+
(\overline{f}\ \overline{F})_L^{down}
\left(
\begin{array}{ccc}
0 & m_L \\
m_R^\dagger & M_F
\end{array} \right) \left(
\begin{array}{c}
f \\
F
\end{array} \right)_R^{down}
+
{\rm h.c.} \ . \eqno(2.2)
$$
In the present model, since we will choose $m_L^\dagger=m_L$ and 
$m_R^\dagger=m_R$ later, what we should do is to diagonalize the 
$6\times 6$ mass matrix 
$$
M=\left(
\begin{array}{ccc}
0 & m_L \\
m_R & M_F
\end{array} \right) \ . \eqno(2.3)
$$
Under the approximation of $M_F\gg m_L, m_R$, 
we obtain a seesaw-type mass matrix form 
$$
M_f\simeq m_L M_F^{-1}m_R \ . \eqno(2.4)
$$
The structures of $m_L$ and $m_R$ are common to quarks and leptons.
The variety of $M_f$ comes from structures of $M_F$ which 
depend on $F=U,D,N$ and $E$.

In the present paper, we assume that 
$$ 
\langle\phi^0_R\rangle_0 \propto \langle\phi^0_L\rangle_0 \ . \eqno(2.5)
$$
i.e., each term in $V(\phi_R)$ takes the coefficient 
which is exactly proportional to the corresponding term in $V(\phi_L)$. 
This assumption means that there is a kind of ``conspiracy" between 
$V(\phi_R)$ and $V(\phi_L)$. 
However, in this paper, we will not go into this problem moreover.

On the other hand, the heavy fermion mass matrices $M_F$ are
given by 
$M_F=y_{0}\left[\langle\Phi_0\rangle_0  {\bf 1}+(y^F_{X}/y_0)
\langle\Phi_X\rangle_0 \right]$,
where $\langle\Phi_0\rangle_0=V_0$, 
$\langle\Phi_X\rangle_0= V^F_{X} X$ and 
$$
{\bf 1} \equiv \left(
\begin{array}{ccc}
1 & 0 & 0 \\
0 & 1 & 0 \\
0 & 0 & 1 
\end{array} \right) \ \ \  {\rm and} \ \ \ 
X \equiv \frac{1}{3} \left(
\begin{array}{ccc}
1 & 1 & 1 \\
1 & 1 & 1 \\
1 & 1 & 1 
\end{array} \right) \ .  \eqno(2.6)
$$
Note that the U(3)-family symmetry is badly broken by 
 $\langle\Phi_X\rangle_0$.
The democratic term $X$ in $M_F$ may be understood, 
for example, by a permutation group S$_3$ [8].
However, in the present stage, since our interest is 
focused on the light Higgs boson $\phi_L$, 
we do not touch the origin of the democratic term $X$. 
Anyhow, we assume that the $M_F$ is given by 
$$ 
M_F = m_0 K_f \left( {\bf 1}+ 3 b_f X \right)
\ ,  \eqno(2.7)
$$
where $m_0K_f=y_0 V_0$ and $3b_f=(y_X^F/y_0)(V_X/V_0)$.
Since  the parameter $K_f$ is given by 
$K_f=y_0 \langle\Phi_0\rangle_0$,  it is independent of 
$f=u,d,e,\nu$, i.e., $K_e=K_u=K_d$ at $\mu=m_0 K$. 
However, for numerical evaluations, we will treat$K_f$ 
effectively as $K_e\neq K_u \simeq K_d$, 
because of the evolution from $\mu\sim m_0 K$ to $\mu\sim M_W$. 

The variety of the quark and lepton mass matrices 
$M_f$ essentially originates in the parameter $b_f$. 
Since we take $v_L\equiv\langle\phi^0_L\rangle_0$ 
and $v_R\equiv\langle\phi^0_R\rangle_0$ such as they are Hermitian,
a $CP$ violation phase can be included only 
in the heavy fermion mass matrix $M_F$ 
(i.e., in the parameter $b_f$).
For the charged leptons, considering the phenomenological 
success of the relation (1.8) [i.e., (1.4)], we put $b_e=0$. 
Therefore, the mass matrix of the heavy leptons $M_E$ is 
Hermitian, and $CP$ violation does not manifest in the charged lepton 
sector.

The phenomenological study of the quark mass spectrum and family-mixings 
based on the seesaw-type mass matrix (1.1) with 
$M_F$ of the form (2.7) has been done by one of the authors (Y.K.) 
and Fusaoka [9].
They  have found that the seesaw-type mass matrix with 
$M_F$ of the form (2.7) can provide an explanation 
why $m_t\gg m_b$, while $m_u\sim m_d$, by taking 
$b_u= -1/3$  ($\beta_u=0$) and $b_d\simeq - e^{i\beta_d}$ 
($\beta_d \simeq - 20^\circ$), but with keeping $K_u=K_d$:
The inverse of the matrix [(unit matrix)$+$(democratic-type matrix)], 
${\bf 1}+3 b_f X$, also take a form 
[(unit matrix)$+$(democratic-type matrix)], 
${\bf 1} + 3 a_f X$, with 
$a_f  =- b_f /(1+3b_f )$; 
The enhancement $m_t/m_b\gg 1$ comes from 
$|a_f|\rightarrow \infty$ in the limit of $b_f\rightarrow -1/3$, while
$m_u\sim m_d$ comes from the feature 
that the democratic-type mass matrix [10] can provide a large mass 
only to the third family, i.e., the effect of $|a_u|\rightarrow\infty$ 
contributes mainly to $m_t$.
Of course, 
by adjusting the parameter $\beta_d$, they [9]  have also obtained reasonable 
KM matrix parameters as well as up- and down-quark masses.
Here, we do not repeat their numerical results.

Since the VEV of $\phi^0_L$ is small compared with those of other Higgs 
fields $\phi^0_R$, $\Phi_0$ and $\Phi_X$, i.e., 
(Tr$\langle\phi_L\rangle_0^2)^{1/2}\sim 10^2$ GeV, 
we expect some observable effects of the physical Higgs bosons $\phi_L$ 
in the low energy ($10^2-10^3$ GeV) experiments.
The purpose of the present paper is to investigate the physics of the 
U(3)$_{family}$ nonet Higgs bosons $\phi_L$ from the phenomenological 
point of view.
In the next section, we investigate a possible form of $V(\phi_L)$ 
which derives the relation (1.8) (therefore  the charged lepton 
mass relation (1.4)).
However, we will not touch what mathematical requirements can provide such 
a potential form. 
The purpose of the present paper is to study  
masses of the physical Higgs bosons $\phi_L$ 
and their interactions with gauge bosons and fermions 
when the fields $\phi_L$ are described by such a potential
which can lead to the relation (1.8).

\vglue.3in

\noindent{\bf 3. Higgs potential $V(\phi_L)$ and ^^ ^^ nonet" ansatz}

\vglue.1in

What is of great interest to us is a potential form of $\phi_L$, 
$V(\phi_L)$.
Hereafter, we will omit the index $L$ and  simply write $\phi_L$ as $\phi$.
We do not consider mixings among Higgs scalar fields with hierarchically 
different VEV's, i.e., among $\phi_L$, $\phi_R$ and $\Phi_{0,X}$.
Then, the potential $V(\phi)$ is given by
$$
V(\phi)=V_{nonet} + V_{Oct\cdot Singl} + V_{SB} \ , \eqno(3.1)
$$
where 
$V_{nonet}$ is a part of $V(\phi)$ which satisfies a ``nonet" ansatz 
stated below, $V_{Oct\cdot Singl}$ is a part which violates the ``nonet" 
ansatz, and $V_{SB}$ is a term which breaks U(3)$_{family}$ explicitly. 

The ``nonet" ansatz is as follows: the octet component $\phi_{oct}$ 
and singlet component $\phi_s$ of the Higgs scalar fields $\phi_L$ 
($\phi_R$) always appear with the combination of (1.6) in the 
Lagrangian. 
Under the ``nonet" ansatz, the SU(2)$_L$ invariant (and also U(3)$_{family}$ 
invariant) potential $V_{nonet}$ is given by
$$
V_{nonet}= \mu^2 {\rm Tr}(\overline{\phi}\phi)
+\frac{1}{2}\lambda_1 
\sum_{i,j}\sum_{k,l}(\overline{\phi}_i^j \phi_j^i)
(\overline{\phi}_k^l \phi_l^k) 
$$
$$
+\frac{1}{2}\lambda_2 
\sum_{i,j}\sum_{k,l}(\overline{\phi}_i^j \phi_k^l)
(\overline{\phi}_l^k \phi_j^i) 
+\frac{1}{2}\lambda_3 
\sum_{i,j}\sum_{k,l}
(\overline{\phi}_i^j \phi_k^l)(\overline{\phi}_j^i \phi_l^k) 
\ , \eqno(3.2)
$$ 
where $(\overline{\phi} \phi)=\phi^-\phi^+ 
+\overline{\phi}^0 \phi^0$.
Here, for simplicity, we have taken only U(3)$_{family}$ 
singlet terms in which each two of four fields can make 
U(3)$_{family}$ singlets. A more general case, which includes terms 
$\sum_{i,j,k,l}(\overline{\phi}_i^j\phi_j^k)(\overline{\phi}_k^l
\phi_l^i)$ 
and so on, is given in Appendix A.

In addition to $V_{nonet}$ which satisfies the nonet ansatz, 
we consider terms the following interaction terms between octet- and 
singlet-components 
$V_{Oct\cdot Singl}$ which break the nonet ansatz:
$$
V_{Oct\cdot Singl}=
\eta_1(\overline{\phi}_s\phi_s){\rm Tr}(\overline{\phi}_{oct} \phi_{oct})
+ \eta_2\sum_{i,j}\left( \overline{\phi}_s(\phi_{oct})_i^j\right)
\left( (\overline{\phi}_{oct})_j^i \phi_s\right)
$$
$$
+ \eta_3\sum_{i,j}\left( \overline{\phi}_s(\phi_{oct})_i^j\right)
\left( \overline{\phi}_s (\phi_{oct})_j^i\right)
+ \eta_3^*\sum_{i,j}\left( (\overline{\phi}_{oct})_i^j \phi_s\right)
\left( (\overline{\phi}_{oct})_j^i \phi_s\right) \ . \eqno(3.3)
$$
Note that the both potential terms $V_{nonet}+V_{Oct\cdot Singl}$ are
invariant under SU(3)$_{family}$ symmetry and 
the exchange $\phi_{oct}\leftrightarrow 
{\bf 1}(\phi_s/\sqrt{3})$.

As stated later, the potential which consists only of $V_{nonet}$ and 
$V_{Oct\cdot Singl}$ 
cannot fix each value $v_i$ of the VEV's $\langle\phi^0\rangle_0=v=
{\rm diag}(v_1, v_2, v_3)$ completely, although we can derive that 
the VEV's $v$ should satisfy the relation (1.8).
In order to fix three values of $v_i$ completely, we will add to 
an explicitly U(3)$_{family}$ symmetry breaking term $V_{SB}$. 
We consider that 
gauge symmetries are exact symmetries in the original Hamiltonian, 
so that those are broken only spontaneously,
while  global symmetries are phenomenological and approximate 
symmetries, so that the symmetries may be broken explicitly.

For a time, we neglect the term $V_{SB}$ in (3.1).
For $\mu^2<0$, conditions for minimizing the potential (3.1) are as follows:
$$
\left[\mu^2+(\lambda_1+\lambda_2){\rm Tr}(v^{\dagger}v)\right]v_s^*
+ \lambda_3{\rm Tr}(v^{\dagger}v^{\dagger})v_s
$$
$$
+ (\eta_1+\eta_2){\rm Tr}(v^{\dagger}_{oct}v_{oct})v_s^*
+ 2\eta_3^*{\rm Tr}(v^{\dagger}_{oct}v^{\dagger}_{oct})v_s=0 \ , \eqno(3.4)
$$
$$
\left[\mu^2+(\lambda_1+\lambda_2){\rm Tr}(v^{\dagger}v)\right]
v_{oct}
+ \lambda_3{\rm Tr}(v^{\dagger}v^{\dagger})v_{oct}
+ (\eta_1+\eta_2)v_s^*v_s^*v^{\dagger}_{oct}
+ 2\eta_3v_s^*v_s^*v_{oct}=0 \ , \eqno(3.5)
$$
and the similar equations with $v\leftrightarrow v^\dagger$ 
($v_{oct}\leftrightarrow v^\dagger_{oct}$ and 
$v_s\leftrightarrow v_s^\dagger$). 
For simplicity, we consider the case $\eta_3^*=\eta_3$, so that 
$v^\dagger=v$.
Then, 
we can readily obtain the desirable relation
$$
v_s^2={\rm Tr}(v_{oct}^2)=\frac{-\mu^2}
{2(\lambda_1+\lambda_2+\lambda_3)+\eta_1+\eta_2+2\eta_3} \ , \eqno(3.6)
$$
which leads to the relation (1.8).

Note that the exact nonet form (1.6) is not always essential to 
provide the relation (1.8).
In the modified potential 
$$
V_{nonet}= \mu_s^2(\overline{\phi}_s\phi_s) + 
\mu^2_{oct} {\rm Tr}(\overline{\phi}_{oct}\phi_{oct})
+\frac{1}{2}\lambda_1 \sum_{i,j} \sum_{k,l}
(\overline{\phi}_i^j \phi_j^i)(\overline{\phi}_k^l \phi_l^k) +\cdots 
\ , \eqno(3.7)
$$
where
$$
\phi=\phi_{oct}+\frac{k}{\sqrt{3}}\phi_s\; {\bf 1}\ ,\eqno(3.8)
$$
we can also obtain the desirable relation (1.8) when the coefficient $k$ 
satisfies 
$$
    k^2=\mu_s^2/\mu_{oct}^2 \ , \eqno(3.9)
$$
because the conditions for minimizing (3.7) leads to 
$$
k^2 v_s^2={\rm Tr}(v_{oct}^2)=\frac{-\mu^2_s}
{2(\lambda_1+\lambda_2+\lambda_3)+\eta_1+\eta_2+2\eta_3} \ . \eqno(3.10)
$$
The essential assumption is that the terms $\overline{\phi}_s\phi_s$ 
and $\overline{\phi}_{oct}\phi_{oct}$ 
appear with the same relative weight in $V_{nonet}$ 
and in the Yukawa interactions with fermions.
However, there is no substantial difference between the cases of $k=1$ 
and $k\neq 1$ for evaluation of physical quantities. 
Therefore, hereafter, we will investigate only the case of $k=1$.

So far, we do not have any conditions more than (3.6) 
(therefore (1.8)) for $v$, 
although it is sufficient for deriving charged lepton mass relation (1.4). 
In order to fix each component of $v$, we must add some additional terms
to the potential $V(\phi)$. 

In general, we can choose such a family-basis in which $v$ is given by 
a diagonal form 
$$
      v={\rm diag}(v_1,v_2,v_3) \ . \eqno(3.11)
$$
Then, under the replacement $\phi^0\rightarrow\phi^0 + v$, 
seven components of $\phi^0_{oct}$, i.e., six components $\phi_i^{0j}$
($i\neq j$) and a diagonal component (we denote it as $\phi_y^0$) 
can invariant, although the singlet component $\phi^0_s$ and 
the other diagonal component $\phi^0_x$ which is orthogonal to $\phi^0_y$ 
cannot be invariant:
$$
\begin{array}{ll}
\phi^0_s & \rightarrow\phi^0_s + v_s \ , \\
\phi^0_x & \rightarrow\phi^0_x + v_x \ , \\
\phi^0_y & \rightarrow\phi^0_y \ , \\
(\phi^0)_i^j & \rightarrow(\phi^0)_i^j \ \ (i\neq j) \ . 
\end{array} \eqno(3.12)
$$
Therefore, we add the following U(3)$_{family}$ symmetry breaking terms 
to the potential of $\phi$:
$$
V_{SB}=\xi \left[ (\overline{\phi}_y \phi_y)
+\sum_{i,j}(\overline{\phi}_i^{0_j}\phi_j^{0_i}) \right] \ . 
\eqno(3.13)
$$
We can easily see that the relation (3.6) is unchanged even by 
adding such the explicitly symmetry-breaking terms $V_{SB}$, 
because of (3.12). 
We would like to stress that the explicit U(3)$_{family}$ breaking 
(3.13) is a soft breaking, so that it does not spoil the Yukawa sector.
We consider that the parameter $\xi$ satisfies
$$
\xi + \mu^2 >0 \ , \eqno(3.14)
$$
in order to guarantee $\langle\phi^0_y\rangle_0=0$ and 
$\langle\phi^{0j}_i\rangle_0=0$ ($i\neq j$).

We can rewrite the mass terms 
$\mu^2 {\rm Tr}(\overline{\phi}\phi) + V_{SB}$ into
$$
\mu^2(\overline{\phi}_s\phi_s)
+\mu_{oct}^2 {\rm Tr}(\overline{\phi}_{oct}\phi_{oct}) +V'_{SB} \ , 
\eqno(3.15)
$$
where
$$
V'_{SB}= -\xi (\overline{\phi}_x\phi_x) \ , \eqno(3.16)
$$
and $\mu_{oct}^2=\xi+\mu^2$.
The term $V'_{SB}$ plays a role to fix the axis of 
the SU(3)$_{family}$ breaking.

Here, for convenience of our discussions, we define the parameters $z_i$ as
$$ v_i=v_0 z_i \ , \eqno(3.17) $$
$$ v_0=(v_1^2+v_2^2+v_3^2)^{{1}/{2}} \ , \eqno(3.18) $$
with $ z_1^2+z_2^2+z_3^2=1$. 
Also we define the diagonal components of $\phi_{oct}$, 
$\phi_x$ and $\phi_y$, as
$$
\begin{array}{l}
\phi_x =x_1\phi_1^1+x_2\phi_2^2+x_3\phi_3^3 \ , \\
\phi_y =y_1\phi_1^1+y_2\phi_2^2+y_3\phi_3^3 \ , 
\end{array} \eqno(3.19)
$$
with $\sum_i x_i=0$, $\sum_i y_i=0$, $\sum_i x_i^2=1$, $\sum_i y_i^2=1$ 
and $\sum_i x_i y_i=0$,  
where $v_i=v_s/\sqrt{3}+x_i v_x$.
Then, the coefficients $x_i$ and $y_i$ are given by the following relations:
$$ x_i={\sqrt{2}} z_i -\frac{1}{\sqrt{3}} \ , \eqno(3.20) $$
$$
(y_1,y_2,y_3)=(\frac{x_2-x_3}{\sqrt{3}},\frac{x_3-x_1}{\sqrt{3}},
\frac{x_1-x_2}{\sqrt{3}}) \ . \eqno(3.21)
$$
Some useful formulas for the parameters $z_i$ are given in Appendix B.

Although in the present stage of the model, 
we must add an SU(3)$_{family}$ symmetry breaking 
term $V'_{SB}$ by ``hand", this does not mean that we must 
provide three values $(x_1,x_2,x_3)$. 
The independent parameter of $x_i$ is only one (for example, 
see (B7) in Appendix B).

\vglue.3in

\noindent{\bf 4. Higgs boson masses and interactions}

For convenience, we rewrite the fields $\phi^{\pm}$ and $\phi^{0}$ 
with the fields $\chi^{\pm}$, $\chi^0$, and $H^0$ defined by 
$$
\begin{array}{ccc}
\left( 
\begin{array}{c}
\phi^+ \\
\phi^0 
\end{array} \right)
& = & \frac{1}{\sqrt{2}}
\left( 
\begin{array}{c}
i\sqrt{2} \, \chi^+ \\
H^0-i\chi^0
\end{array} \right)
\end{array}
\ , \eqno(4.1)
$$
Then, the mass terms after the spontaneous symmetry breakdown 
are given by 
$$
V_{mass}=V_{m}({\chi^{\pm}})+V_{m}({\chi^0})+V_{m}({H^0})
\ , \eqno(4.2)
$$
$$
V_{m}({\chi^{\pm}})=
\xi \left[ \sum_{i\neq j}(\chi^-)_i^j(\chi^+)_j^i+\chi_y^-\chi_y^+\right]
$$
$$
+(\lambda_2+\lambda_3)\left[ {\rm Tr}(v\chi^+){\rm Tr}(v\chi^-)
-2v_s^2{\rm Tr}(\chi^-\chi^+)\right]
$$
$$
+(\eta_2+2\eta_3)\left[ v_s\chi_s^-{\rm Tr}(v\chi^+)
+v_s\chi_s^+{\rm Tr}(v\chi^-)
-2v_s^2\chi_s^-\chi_s^+-v_s^2{\rm Tr}(\chi^-\chi^+)\right]
\ , \eqno(4.3)
$$

$$
V_{m}({\chi^0})=
\frac{1}{2} \xi\left[ \sum_{i\neq j}(\chi^0)_i^j(\chi^0)_j^i
+\chi_y^0\chi_y^0\right]
+ \lambda_3 \left\{ [{\rm Tr}(v\chi^0)]^2
-2v_s^2{\rm Tr}(\chi^0\chi^0)\right\}
$$
$$
+2\eta_3\left[ -v_s^2{\rm Tr}(\chi^0\chi^0)
+2v_s\chi_s^0{\rm Tr}(v\chi^0)-2v_s^2\chi_s^0\chi_s^0\right]
\ , \eqno(4.4)
$$
$$
V_{m}({H^0})=
\frac{1}{2} \xi\left[ \sum_{i\neq j}(H^0)_i^j(H^0)_j^i+H_y^0H_y^0\right]
+ (\lambda_1+\lambda_2+\lambda_3)[{\rm Tr}(vH^0)]^2
$$
$$
+2(\eta_1+\eta_2+2\eta_3)v_sH_s^0\left[ {\rm Tr}(vH^0)-v_sH_s^0\right]
\ , \eqno(4.5)
$$
where we have used the relations (3.4) and (3.5).

First, we discuss masses of the charged Higgs bosons 
$\chi^{\pm}$. From (4.3), 
the mass terms $\sum_{i,j}(\chi^-)_i^iM^2_{ij}(\chi^+)_j^j$ 
for the diagonal components of the fields $\chi^{\pm}$ are given by
$$
M_{ij}^2=\xi y_iy_j+(\lambda_2+\lambda_3)(v_iv_j-v_0^2\delta_{ij})+
$$
$$
+\frac{1}{3}(\eta_2+2\eta_3)(v_1+v_2+v_3)\left[(v_i+v_j)-(v_1+v_2+v_3)
(\frac{2}{3}+\delta_{ij})\right] \ , \eqno(4.6)
$$
where, from (3.20) and (3.21), $y_i$ is given by $y_i=\sqrt{2}(v_j-v_k)/
\sqrt{3}v_0$
($(i,j,k)$ are cyclic indexes of $(1,2,3)$).

The mass matrix (4.6) is diagonalized by 
transforming ($\phi_1^1, \phi_2^2, \phi_3^3$) into 
$$
\begin{array}{cccc}
\left( 
\begin{array}{c}
\phi_1 \\
\phi_2 \\
\phi_3 
\end{array} \right)
& = & 
\left( 
\begin{array}{ccc}
z_1 & z_2 & z_3 \\
z_1-\sqrt{\frac{2}{3}} & z_2-\sqrt{\frac{2}{3}} & z_3-\sqrt{\frac{2}{3}} \\
\sqrt{\frac{2}{3}}(z_2-z_3) & \sqrt{\frac{2}{3}}(z_3-z_1) 
& \sqrt{\frac{2}{3}}(z_1-z_2) 
\end{array} \right)
\left( 
\begin{array}{c}
\phi_1^1 \\
\phi_2^2 \\
\phi_3^3
\end{array} \right)
\end{array}
\ , \eqno(4.7)
$$
($\phi=\chi^{\pm}$, $\chi^0$ and $H^0$).  From (3.20) 
and (3.21), we find
$$
\begin{array}{ccc}
\left( 
\begin{array}{c}
\phi_1 \\
\phi_2 \\
\phi_3
\end{array} \right)
& = & \frac{1}{\sqrt{2}}
\left( 
\begin{array}{c}
\phi_x+\phi_s \\
\phi_x-\phi_s \\
\sqrt{2}\,\phi_y
\end{array} \right)
\end{array}
\ . \eqno(4.8)
$$
Then, we obtain the masses of $\chi^\pm$ as follows : 
$$
\begin{array}{l}
m^2(\chi_1^\pm)=0 \ , \\
m^2(\chi_2^\pm)=-(\lambda_2+\lambda_3+\eta_2+2\eta_3)v_0^2 \ , \\
m^2(\chi_3^\pm)=-\left[\lambda_2+\lambda_3+\frac{1}{2}(\eta_2+2\eta_3) -
\overline{\xi}\right]v_0^2 \ , \\
m^2(\chi_i^{{\pm}_j})=-\left[\lambda_2+\lambda_3+\frac{1}{2}
(\eta_2+2\eta_3) - \overline{\xi}\right]v_0^2 \ , 
\end{array} \eqno(4.9)
$$
where 
$$
\overline{\xi}=\xi/v_0^2 \ , \eqno(4.10)
$$ 
and $\chi_i^j$ denotes a boson with $i\neq j$. 

Similarly, by the transformation (4.7), 
we obtain masses of $\chi^0$ and $H^0$ : 
$$
\begin{array}{l}
m^2(\chi_1^0)=0 \ , \\
m^2(\chi_2^0)=-2(\lambda_3+2\eta_3)v_0^2 \ , \\
m^2(\chi_3^0)=-\left[2(\lambda_3+\eta_3)-
\overline{\xi}\right]v_0^2 \ , \\
m^2(\chi_i^{0j})=-\left[2(\lambda_3+\eta_3)-
\overline{\xi}\right]v_0^2 \ , 
\end{array} \eqno(4.11)
$$
and
$$
\begin{array}{l}
m^2(H_1^0)=\left[ 2(\lambda_1+\lambda_2+\lambda_3)+\eta_1+\eta_2+2\eta_3
\right]v_0^2 \ , \\
m^2(H_2^0)=-(\eta_1 + \eta_2+2\eta_3)v_0^2 \ , \\
m^2(H_3^0)=\overline{\xi}v_0^2 \ , \\ 
m^2(H_i^{0j})=\overline{\xi}v_0^2 \ . 
\end{array} \eqno(4.12)
$$

Since $\mu^2< 0$, the positivity of $m^2(H_1^0)$ is 
guaranteed by (3.6), i.e., 
$$
2(\lambda_1+\lambda_2+\lambda_3)+\eta_1+\eta_2+2\eta_3 
=-2\mu^2/v_0^2 > 0 \ . \eqno(4.13)
$$
The positivities of $m^2(H_2^0)$ and $m^2(\chi_2^{\pm}$) require 
$$
2(\lambda_1+\lambda_2+\lambda_3) > -(\eta_1+\eta_2+2\eta_3) > 0 
\ , \eqno(4.14)
$$
and 
$$
2\lambda_1+\lambda_2 +\lambda_3+\eta_1 > 
-(\lambda_2+\lambda_3+\eta_2+2\eta_3) > 0 
\ , \eqno(4.15)
$$
respectively. 
These relations are consistent with the positivity conditions of 
the potential for large values of the fields $\phi$.  From 
(4.13) and (3.14), we obtain a constraint 
$$
2m^2(H^0_3) > m^2(H^0_1) \ . \eqno(4.16)
$$

If we suppose $\overline{\xi}\rightarrow\infty$, the Higgs bosons
$\chi_i^{\pm j}$, $\chi_3^\pm$, $\chi_i^{0 j}$, $\chi_3^0$, $H_i^{0 j}$
and $H_3^0$ decouple from the present model, the physical Higgs bosons 
become only $H_1^0$, $H_2^0$, $\chi_2^0$ and $\chi_2^\pm$, 
so that the model becomes essentially identical with the 
two-Higgs-doublet model [11].

On the other hand, weak boson masses are obtained by calculating 
the kinetic term Tr($D_{\mu}\overline{\phi}D^{\mu}\phi$) 
($D_{\mu}$ is a covariant derivative).  From 
the straightforward calculation, we can check that 
the Higgs bosons which are eaten by weak bosons $W^{\pm}$ and $Z^0$ are 
$\chi_1^{\pm}$ and $\chi_1^0$. The masses of weak bosons 
are given by 
$$
m^2(W^{\pm})=\frac{1}{2}g^2v_0^2 \ , \eqno(4.17)
$$
$$
m^2(Z^0)=\frac{1}{2}g_z^2v_0^2  \ , \eqno(4.18)
$$
where $g_z=g/\cos\theta_W$.
Therefore, the value of $v_0$ is given by 
$$
v_0^2 =\frac{m^2(W^\pm)}{g^2/2}=\frac{1}{4G_F/\sqrt{2}}=(174 \ {\rm GeV})^2
\ . \eqno(4.19)
$$

Since we take interest only in new effects which are caused by the existence 
of $\phi_L$, we neglect mixing of $\phi_L$ with $\phi_R$. 
Since, for a time, we deal only with tree-level physics, we calculate 
the interactions of $\phi_L$ by taking the unitary gauge.

Interactions of $\phi_L$ with gauge bosons are calculated from the kinetic 
term Tr($D_{\mu}\overline{\phi}_{L}D^{\mu}\phi_L$). The results are as 
follows : 
$$
H_{EW}=
+\frac{1}{2}\left( 2gm_W W_{\mu}^-W^{+\mu}
+g_z m_Z Z_{\mu}Z^{\mu}\right) H_1^0
$$
$$
+i\left( eA_{\mu}+\frac{1}{2}g_z\cos2\theta_W Z_{\mu}\right) {\rm Tr}
(\chi^{-}\buildrel\leftrightarrow\over\partial\,^{\mu} \chi^+)
+\frac{1}{2}g_z Z_{\mu}{\rm Tr}
(\chi^{0} \buildrel\leftrightarrow\over\partial\,^{\mu} H^0)
$$
$$
+\left( e^2A_{\mu}A^{\mu}+eg_{z}\cos2\theta_W A_{\mu}Z^{\mu}
+\frac{1}{4}g_z^2\cos^22\theta_W Z_{\mu}Z^{\mu}\right){\rm Tr}(\chi^-\chi^+)
$$
$$
+\frac{1}{8}g_z^2 Z_{\mu}Z^{\mu}[{\rm Tr}(\chi^0\chi^0)+{\rm Tr}(H^0H^0)]
$$
$$
+\frac{1}{4}g^2W_{\mu}^-W^{+\mu}\left[ 2{\rm Tr}(\chi^-\chi^+)
+{\rm Tr}(\chi^0\chi^0)+{\rm Tr}(H^0H^0)\right]
$$
$$
-\frac{1}{2}g \left(eA_{\mu}-g_z \sin^2\theta_W Z_{\mu}\right)
\left\{ W^{+\mu}[{\rm Tr}(\chi^-\chi^0)+i{\rm Tr}(\chi^-H^0)]+{\rm h.c.}
\right\}
\ , \eqno(4.20)
$$
where $g_z=g/\cos\theta_W$ and $\chi_1^{\pm}=\chi_1^0=0$. Note that 
the interactions of the neutral Higgs boson $H_1^0$ with gauge bosons are 
completely identical with those of the neutral Higgs boson $H^0$ in the 
standard model. 

Three-body interactions of $\phi_L$ are calculated from the potential 
(3.1):
$$
H_{\phi\phi\phi}=\sqrt{2}\lambda_1v_0{\rm Tr}(\chi^-\chi^+)H_1^0
+\frac{1}{\sqrt{2}}(\lambda_1+\lambda_2)v_0{\rm Tr}(\chi^0\chi^0+H^0H^0)
H_1^0
$$
$$
-\frac{1}{\sqrt{2}}\lambda_3v_0{\rm Tr}(\chi^0\chi^0-H^0H^0)H_1^0
+\sqrt{2}i\lambda_4{\rm Tr}[(\chi^0v-v\chi^0)\chi^-\chi^+]
$$
$$
+\frac{1}{2\sqrt{2}}\lambda_5\{{\rm Tr}\left[(H^0v+vH^0)(\chi^-\chi^{+}+\chi^+
\chi^-)\right]
-i{\rm Tr}\left[(\chi^0v-v\chi^0)(\chi^-\chi^+-\chi^+\chi^-)\right]
$$
$$
-2{\rm Tr}[v(\chi^+H^0\chi^-+\chi^-H^0\chi^+)]\}
$$
$$
+\frac{1}{2\sqrt{2}}\eta_1v_0\left[ (H_1^0-H_2^0){\rm Tr}(2\chi^-\chi^{+}
+\chi^0\chi^0+H^0H^0)\right.
$$
$$\left.
+H_2^0(2\chi_2^-\chi_2^{+}+H_1^0H_1^0+H_2^0H_2^0-2H_1^0H_2^0)\right]
$$
$$
+\frac{1}{\sqrt{2}}\eta_2v_0\{-\chi_2^-{\rm Tr}[\chi^+(H^0+i\chi^0)]
+{\rm h.c.}
$$
$$
+(H_1^0-H_2^0)[{\rm Tr}(H^0H^0+\chi^0\chi^0)-2\chi_2^-\chi_2^+]
$$
$$
+H_2^0(\chi_2^0\chi_2^0+H_1^0H_1^0+H_2^0H_2^0-2H_1^0H_2^0)]\}
$$
$$
+\frac{1}{\sqrt{2}}\eta_3\{-\chi_2^-{\rm Tr}[\chi^+(H^0-i\chi^0)]
+{\rm h.c.}-2\chi_2^0{\rm Tr}(\chi^0H^0)
$$
$$
+(H_1^0-H_2^0)[{\rm Tr}(H^0H^0-\chi^0\chi^0)-2\chi_2^-\chi_2^+]
$$
$$
+H_2^0(H_1^0-H_2^0)^2-(2H_1^0-H_2^0)\chi_2^0\chi_2^0\}
\ . \eqno(4.21)
$$
As we discuss in Sect.5 and Sect.6, we consider that only the Higgs 
boson $H_1^0$ is light compared with the other Higgs bosons 
whose masses are of the order of TeV. 
The interaction (4.21) states the absence of the decays of these 
Heavy Higgs bosons into two $H_1^2$ states. 
Of course, the lightest Higgs boson $H_1^0$ cannot decay into 
multi-Higgs-boson states.
The dominant decay modes of our Higgs bosons are those into 
two fermion states.

\vglue.3in

\noindent{\bf 5. Effective fermion-Higgs interactions and 
$K_L\rightarrow \mu^{\pm}e^{\mp}$ decay}

Our Higgs particles $\phi_L$ do not have interactions with light fermions 
$f$ at tree level, and they can couple only between light fermions $f$ and 
heavy fermions $F$. 
However, since the fermion mass matrix (2.3) is diagonalized, the 
physical fermion states (mass eigenstates) are mixed states of $f$ and $F$. 
The physical fermion states are given by 
$$
\begin{array}{ccccc}
\left(
\begin{array}{c}
f_{L}^{phys} \\
F_{L}^{phys}
\end{array} \right)
& = & 
\left(
\begin{array}{cc}
U_{L}^f & 0 \\
0 & U_{L}^F
\end{array} \right)
& U_{L}^{(6\times 6)} & 
\left(
\begin{array}{c}
f_{L} \\
F_{L}
\end{array} \right)
\end{array}
\ , \eqno(5.1)
$$
(and the similar relation with $L\rightarrow R$), 
where $6\times6$ unitary matrices $U_{L}^{(6\times 6)}$ and 
$U_{R}^{(6\times 6)}$ diagonalize the 
$6\times6$ mass matrix (2.3) as 
$$
U_L^{(6\times 6)}
\left(
\begin{array}{cc}
0 & m_L \\
m_R & M_F
\end{array} \right)
U_R^{(6\times 6)\dagger}
=
\left(
\begin{array}{cc}
M_f & 0 \\
0 & M_F^{'}
\end{array} \right)
\ , \eqno(5.2)
$$
and $3\times3$ unitary matrices $U_{L/R}^f$ and $U_{L/R}^F$ 
diagonalize the $3\times3$ mass matrices $M_f$ and $M_F^{'}$ 
in (5.2) into the diagonal matrices $D_f$ and $D_F$ as 
$$
U_L^fM_fU_R^{f\dagger}=D_f \ , \ \ \ \ \ \ 
U_L^F M_F^{'}U_R^{F\dagger}=D_F \ , \eqno(5.3)
$$
respectively.

In the ^^ ^^ seesaw" approximation ($m_L, m_R, \ll M_F$), the 
$6\times6$ unitary matrices 
$U_{L/R}^{(6\times 6)}$ are given by 
$$
U_L^{(6\times 6)} \simeq 
\left(
\begin{array}{cc}
{\bf 1} & -m_LM_F^{-1} \\
M_F^{\dagger-1}m_L & {\bf 1}
\end{array} \right)
\ , \eqno(5.4)
$$
$$
U_R^{(6\times 6)} \simeq 
\left(
\begin{array}{cc}
{\bf 1} & -m_R M_F^{\dagger-1} \\
M_F^{-1} m_R & {\bf 1}
\end{array} \right)
\ , \eqno(5.5)
$$
where $m_L$ and $m_R$ are Hermitian in the present model. 
Therefore, an interaction vertex with the fermions $(f,F)$ 
$$
\left(
\begin{array}{cc}
\Gamma_{11} & \Gamma_{12} \\
\Gamma_{21} & \Gamma_{22}
\end{array} \right)
\ , \eqno(5.6)
$$
is also transformed into 
$$
\left(
\begin{array}{cc}
\Gamma_{11}^{phys} & \Gamma_{12}^{phys} \\
\Gamma_{21}^{phys} & \Gamma_{22}^{phys}
\end{array} \right)
\ , \eqno(5.7)
$$
where
$$
\Gamma_{11}^{phys}\simeq U_L^f\left[ \Gamma_{11}
-\Gamma_{12}M_F^{-1}m_R-m_LM_F^{-1}\Gamma_{21}
+m_LM_F^{-1}\Gamma_{22}M_F^{-1}m_R\right] U_R^{f\dagger}
$$
$$
\simeq U_L^f\left[\Gamma_{11}+\Gamma_{12}m_L^{-1}M_f
+M_f m_R^{-1} \Gamma_{21}
+M_fm_R^{-1}\Gamma_{22}m_L^{-1}M_f\right] U_R^{f\dagger}
\ , \eqno(5.8)
$$
and so on. In (5.8), we have used the relation 
$$
M_f\simeq -m_LM_F^{-1}m_R
\ . \eqno(5.9)
$$

For the interactions of $\phi_L$ with fermions, the vertex (5.6) 
is given by $\Gamma_{12}=y_f\phi_L$ and 
$\Gamma_{11}=\Gamma_{12}=\Gamma_{22}=0$, 
so that we obtain 
$$
\Gamma_{11}^{phys}\simeq U_L^fy_f\phi_Lm_L^{-1} M_f U_R^{f\dagger}
=U_L^f\phi_L\langle\phi_L\rangle_0^{-1} U_L^{f\dagger}D_f
\ , \eqno(5.10)
$$
where we have used the relations $m_L=y_f\langle\phi_L\rangle_0$ and 
$U_L^f M_f U_R^{f\dagger}=D_f$.

For charged leptons $f=e$, since $U_L^e=U_R^e={\bf 1}$, 
$D_e={\rm diag}(m_e, m_{\mu}, m_{\tau})$ 
and $\langle\phi_L\rangle_0^{-1}=v^{-1}=
{\rm diag}(v_1^{-1}, v_2^{-1}, v_3^{-1})$, 
the fields ($\phi_L)_i^j$ couple to 
($\overline{e}_{iL}^{phys}e_{jR}^{phys}$) 
with the effective coupling 
$$
\kappa_j\equiv \frac{m_j^e}{v_j}
=\frac{1}{v_0}\sqrt{m_j^e(m_{\tau}+m_{\mu}+m_e)}
=\frac{m_\tau+m_\mu+m_e}{v_0} z_j \ . \eqno(5.11)
$$

For example, for the interaction of $(H^0-i\chi^0)/\sqrt{2}$, 
we obtain 
$$
\frac{1}{\sqrt{2}}\kappa_j(\overline{e}_L^ie_{Rj})(H^0-i\chi^0)_i^j
+\frac{1}{\sqrt{2}}\kappa_i(\overline{e}_R^ie_{Lj})(H^0+i\chi^0)_i^j
$$
$$
=\frac{1}{2\sqrt{2}}\left\{(\overline{e}^i(a_{ij}-b_{ij}\gamma_5)e_j)
(H^0)_i^j
+i(\overline{e}^i(b_{ij}-a_{ij}\gamma_5)e_j)(\chi^0)_i^j\right\}
\ , \eqno(5.12)
$$
where
$$
a_{ij}=\kappa_i+\kappa_j \ , \ \ \ \ b_{ij}=\kappa_i-\kappa_j \ . \eqno(5.13)
$$
Therefore, in the pure leptonic modes, the exchange of $\phi_L$ 
cannot cause family-number non-conservation. 
On the other hand, in quark sector, since $U_{L/R}^q\neq1$, the family-number 
non-conservation is, in general, caused by the exchange of $\phi_L$. 

Note that even in the limit of $U_L^f=1$, the Higgs bosons 
$(H^0)_i^j$ ($i\neq j$) can sensitively affect 
rare decay modes $K_L\rightarrow e^\mp \mu^\pm $, 
$K^+\rightarrow \pi^+e^- \mu^+$, 
$D^+\rightarrow \pi^+\mu^- e^+$, $B^+\rightarrow \pi^+ e^- \tau^+$, 
$B^+\rightarrow K^+\mu^- \tau^+$, and so on.
The most rigorous bound of the mass of the Higgs bosons $(H^0)_i^j$ 
($(\chi^0)_i^j$ ($i\neq j$) comes from the rare decay 
$K_L\rightarrow e^\mp \mu^\pm $.
The effective Hamiltonian for the decay $s\rightarrow d+e^++\mu^-$ is 
given by 
$$ H_{eff}=\frac{1}{m^2_H}(\overline{d}(a_{sd}+b_{sd}\gamma_5)s)
(\overline{\mu}(a_{\mu e}-b_{\mu e}\gamma_5)e) 
$$
$$ +\frac{1}{m^2_\chi}(\overline{d}(b_{sd}+a_{sd}\gamma_5)s)
(\overline{\mu}(b_{\mu e}-a_{\mu e}\gamma_5)e) \ , \eqno(5.14)
$$
where
$$
a_{sd}\simeq \frac{1}{2\sqrt{2}}\left(\frac{m_s}{v_2}+\frac{m_d}{v_1}\right)
\ , \ \ \ 
b_{sd}\simeq \frac{1}{2\sqrt{2}}\left(\frac{m_s}{v_2}-\frac{m_d}{v_1}\right)
\ , \eqno(5.15)
$$
$$
a_{\mu e}= \frac{1}{2\sqrt{2}}\left(\frac{m_\mu}{v_2}+\frac{m_e}{v_1}\right)
\ , \ \ \ 
b_{\mu e}= \frac{1}{2\sqrt{2}}\left(\frac{m_\mu}{v_2}-\frac{m_e}{v_1}\right)
\ , \eqno(5.16)
$$
and we have used $U_d\simeq {\bf 1}$.
By using the relation
$$ 
\langle 0|(\overline{d}\gamma_5 s)|\overline{K}^0(p)\rangle
=\frac{-i}{m_s+m_d} f_K m_K^2 \ , \eqno(5.17)
$$
we obtain the decay amplitude
$$
A(K^0\rightarrow e^- \mu^+) 
= \frac{f_K m_K^2}{m_s+m_d} \left[ \frac{b_{sd}}{m_H^2}
(\overline{\mu}(a_{\mu e}-b_{\mu e}\gamma_5)e) 
 +  \frac{a_{sd}}{m_\chi^2}
(\overline{\mu}(b_{\mu e}-a_{\mu e}\gamma_5)e) \right] \ , \eqno(5.18)
$$
so that 
$$
\Gamma (K_L\rightarrow e^\pm \mu^\mp)\simeq \frac{m_K}{256\pi}
\left(\frac{f_K m_K^2}{m_s+m_d}\right)^2
\left(\frac{1}{m_H^2}+\frac{1}{m_\chi^2}\right)^2
\left(\frac{m_s}{v_2}\right)^2 \left(\frac{m_\mu}{v_2}\right)^2 \ , \
\eqno(5.19)
$$
where we have used the approximation for $(m_d/v_1)^2\ll (m_s/v_2)^2$ and 
$(m_e/v_1)^2\ll (m_\mu/v_2)^2$.
The experimental lower bound [6] 
$B(K_L\rightarrow e^\pm \mu^\mp) < 3.3\times 10^{-11}$ puts the constraint
$$
\left(\frac{1}{m_H^2}+\frac{1}{m_\chi^2}\right)^{-1}
> (1.69 \ {\rm TeV})^2 \ . \eqno(5.20)
$$
If $m_H\sim m_\chi$, (5.20) leads to the lower bound 
$$
m(H_3) \simeq m(\chi_3) > 2.4 \ {\rm TeV} \ . \eqno(5.21)
$$
Thus, Higgs scalar masses are expected to be a few TeV region.

\vglue.3in

\noindent
{\bf 6. Constraints on the Higgs boson masses from $P^0 \overline{P}^0$ 
mixings}

As stated in the previous section, 
since $U_L^q\neq1$ in quark sector,
flavor changing neutral currents (FCNC), in general, appear by exchanging 
the Higgs bosons $\phi_2$, $\phi_3$ and $\phi_i^j$ $(i\neq j)$ 
($\phi=H^0$ and $\chi^0$), and they can sensitively contribute to the 
$K^0\overline{K}^0$, $D^0\overline{D}^0$ and $B^0\overline{B}^0$ 
mixings. In this section, we study the magnitudes of FCNC in details. 

Note that as far as the physical Higgs boson $H_1^0$ is 
concerned, the interaction with quarks $q_i$ is still diagonal type 
($\overline{q}_iq_i)H_1^0$, because $H_1^0=\sum_iv_i(H^0)_i^i/v_0$, 
so that 
$$
\frac{1}{\sqrt{2}}
\frac{1}{v_0}(\overline{q}_LU_L^q v \langle\phi_L\rangle_0^{-1} 
U_L^{q\dagger}D_qq_R)H_1^0 + {\rm h.c.}
= \frac{1}{\sqrt{2}}
\sum_i\frac{m_i^q}{v_0}(\overline{q}_iq_i)H_1^0
\ . \eqno(6.1)
$$
Therefore, the interactions of $H_1^0$ with quarks are identical with 
those of the physical neutral Higgs boson $H_{SM}^0$ in the standard model. 
Recall that the electroweak interactions of $H_1^0$ are also identical 
with those of $H_{SM}^0$. 
As far as $H_1^0$ is concerned, we cannot distinguish from the 
standard-model Higgs boson $H_{SM}^0$ experimentally.

The fermion-Higgs boson interactions are, from (5.10),  given by 
$$
H_{\overline{f}f\phi}=
(\overline{f}_LU\phi v^{-1}U^\dagger Df_R) + {\rm h.c.}
$$
$$
=\sum_{i,j}(\overline{f}_{iL}f_{jR})\left( 
\sum_{k\neq l}\phi_k^l\frac{m_j}{v_l}U_i^kU_j^{l*}
+\sum_k\phi_k^k\frac{m_j}{v_k}U_i^kU_j^{k*} \right) 
+ {\rm h.c.} \ , \eqno(6.2)
$$
where $U\equiv U_L^f$, $\phi\equiv\phi^0=(H^0-i\chi^0)/\sqrt{2}$, 
$D\equiv D_f={\rm diag}(m_1^f, m_2^f, m_3^f$) and $f=u,d$.

The interactions with $\phi_k^l$ $(k\neq l)$ are rewritten as follows: 
$$
\frac{1}{2\sqrt{2}}\sum_{i,j}\sum_{k\neq l}\overline{f}_i
\left[ (A_{ij}^{kl}-B_{ij}^{kl}\gamma_5)(H^0)_k^l + 
i(B_{ij}^{kl}-A_{ij}^{kl}\gamma_5)(\chi^0)_k^l \right]f_j \ , \eqno(6.3)
$$
where
$$
A_{ij}^{kl} = \frac{1}{2}\left(\frac{m_i}{v_k}+\frac{m_j}{v_l}\right)
U_i^kU_j^{l*} \ , \ \ \ 
B_{ij}^{kl} = \frac{1}{2}\left(\frac{m_i}{v_k}-\frac{m_j}{v_l}\right)
U_i^kU_j^{l*} \ . \eqno(6.4)
$$

For the interactions with $\phi_k^k$, by using the expression
$$
\phi_k^k = z_k\phi_1 + \left( z_k-\sqrt{\frac{2}{3}}\right) \phi_2 + 
\sqrt{\frac{2}{3}}(z_l-z_m)\phi_3 \ , \eqno(6.5)
$$
where $(k,l,m)$ are cyclic indices of (1,2,3), we can obtain 
$$
\sum_{i,j}\frac{m_j}{v_0}(\overline{f}_{iL}f_{jR})
\left[ \delta_i^j(\phi_1+\phi_2)-
\sqrt{\frac{2}{3}}\phi_2\sum_k\frac{1}{z_k}U_i^kU_j^{k*}
+ \sqrt{\frac{2}{3}}\phi_3\sum_k\frac{z_l-z_m}{z_k}U_i^kU_j^{k*}
\right] + {\rm h.c.}
$$
$$
=\frac{1}{\sqrt{2}}\sum_i\frac{m_i}{v_0}\left[(
\overline{f}_if_i)(H_1^0+H_2^0)-i(\overline{f}_i\gamma_5f_i)\chi_2^0
\right]
$$
$$
-\frac{1}{\sqrt{3}}\sum_{i,j}\left[(\overline{f}_i(a_{ij}-b_{ij}
\gamma_5)f_j)H_2^0+i(\overline{f}_i(b_{ij}-a_{ij}\gamma_5)f_j)\chi_2^0
\right]
\sum_k\frac{1}{z_k}U_i^kU_j^{k*}
$$
$$
+\frac{1}{\sqrt{3}}\sum_{i,j}\left[(\overline{f}_i(a_{ij}-b_{ij}
\gamma_5)f_j)H_3^0+i(\overline{f}_i(b_{ij}-a_{ij}\gamma_5)f_j)
\chi_3^0 \right]
\sum_k\frac{z_l-z_m}{z_k}U_i^kU_j^{k*} \ , \eqno(6.6)
$$
where
$$
a_{ij}=\frac{1}{2}\left(\frac{m_i}{v_0}+\frac{m_j}{v_0}\right) \ , \ \ \ 
b_{ij}=\frac{1}{2}\left(\frac{m_i}{v_0}-\frac{m_j}{v_0}\right) \ . \eqno(6.7)
$$

The effective four-Fermi interactions of FCNC are given by 
$$
H_{FCNC} = \frac{1}{3}\sum_{i\neq j}
\left[\frac{1}{m^2(H_2^0)}\left(\overline{f}_i
(a_{ij}-b_{ij}\gamma_5)f_j\right)^2 \right. $$
$$\left.-\frac{1}{m^2(\chi_2^0)}\left(\overline{f}_i(
b_{ij}-a_{ij}\gamma_5)f_j\right)^2
\right]\left(\sum_k\frac{1}{z_k}U_i^kU_j^{k*}\right)^2
$$
$$
+\frac{1}{3}\sum_{i\neq j}\left[
\frac{1}{m_2(H_3^0)}\left(\overline{f}_i(a_{ij}-
b_{ij}\gamma_5)f_j\right)^2 \right. $$
$$\left.-\frac{1}{m_2(\chi_3^0)}\left(\overline{f}_i(
b_{ij}-a_{ij}\gamma_5)f_j\right)^2 \right]
\left(\sum_k\frac{z_l-z_m}{z_k}U_i^kU_j^{k*}\right)^2
$$
$$
+\frac{1}{2}\sum_{i\neq j}\sum_{k\neq l}\left[
\frac{1}{m^2(H_k^l)}(\overline{f}_i(A_{ij}^{kl}-B_{ij}^{kl}\gamma_5)f_j)
(\overline{f}_i(A_{ij}^{lk}-B_{ij}^{lk}\gamma_5)f_j)
\right.
$$
$$
\left.
-\frac{1}{m^2(\chi_k^l)}(
\overline{f}_i(
B_{ij}^{kl}-A_{ij}^{kl}\gamma_5)f_j)(
\overline{f}_i(B_{ij}^{lk}-A_{ij}^{lk}
\gamma_5)f_j) \right] \ .
\eqno(6.8)
$$

If we suppose the case $\overline{\xi}\rightarrow\infty$, we can 
neglect the contributions from $\phi_3$ and $\phi_k^l$ $(k\neq l)$ to the 
$K^0\overline{K}^0$ mixing and so on, since these Higgs bosons 
decouple from the low energy effective theory. 
However, even then, the contributions form $\phi_2$ still 
remain.  From the experimental values of $K_L$-$K_S$ 
and $D_1^0$-$D_2^0$ mass differences, the masses of $H_2^0$ and 
$\chi_2^0$ must be large than $10^5$ GeV. 
Considering from (4.14), (4.15) and $v_0=174$ GeV, 
it is unlikely that $H_2^0$ and 
$\chi_2^0$ have such large masses as far as 
we suppose that the coupling constants 
$\lambda_3$, $\eta_1$, $\eta_2$ and $\eta_3$ are of the order 
of one or less than it.

Note that the contributions form $\chi^0$ have the opposite signs 
to that from $H^0$. 
If we suppose $m^2(H^0_k)=m^2(\chi^0_k)\equiv m^2_{Hk}$ $(k=2,3)$, 
which means 
$$
\eta_1+\eta_2=0 \ , \eqno(6.9)
$$
$$
\lambda_3+\eta_3=0 \ , \eqno(6.10)
$$
the contributions of $H_{FCNC}$ are considerably reduced:
$$
H_{FCNC}=\frac{1}{3}
\sum_{i\neq j}\left[
\frac{1}{m^2(H_2^0)}
\left(\sum_k\frac{1}{z_k}U_i^kU_j^{k*}\right)^2
+\frac{1}{m^2(H_3^0)}\left(
\sum_k\frac{z_l-z_m}{z_k}U_i^kU_j^{k*}\right)^2 \right.
$$
$$\left. - 
\frac{3}{2}\frac{1}{m^2(H_i^j)}\sum_k\left(
\frac{1}{z_k}U_i^kU_j^{k*}\right)^2 \right]
\left[
(\overline{f}_if_j)^2-(
\overline{f}_i\gamma_5f_j)^2
\right]
$$
$$
=\frac{1}{3}\left(
\frac{1}{m^2_{H2}}-\frac{1}{m^2_{H3}}\right)
\sum_{i\neq j}\frac{m_im_j}{v_0^2}
\sum_k \left(
\frac{1}{z_k^2}+\frac{z_k-z_l-z_m}{z_1z_2z_3}\right) 
(U_i^kU_j^{k*})^2
$$
$$
\times 
\left[(\overline{f}_if_j)^2-(\overline{f}_i\gamma_5f_j)^2\right] 
\ , \eqno(6.11)
$$
where we have used the relations (B1)--(B4) given in Appendix B and 
$$
U_i^lU_j^{l*}U_i^mU_j^{m*} = 
\frac{1}{2}\left[
(U_i^kU_j^{k*})^2-(U_i^lU_j^{l*})^2-
(U_i^mU_j^{m*})^2\right] \ . \eqno(6.12)
$$

The constraint (6.9) rewrites the $\eta_1$- and $\eta_2$-terms in 
the potential (3.3) into
$$
\eta_1\sum_{i,j}\left[
\phi_s^+(\phi_{oct}^0)_i^j-(\phi_{oct}^+)_i^j\phi_s^0 \right]
\left[\phi_s^-(\overline{\phi}_{oct}^0)_j^i-(\phi_{oct}^-)_j^i
\overline{\phi}_s^0 \right]
\ , \eqno(6.13)
$$
where ($\phi^+\phi^0-\phi^0\phi^+$) and ($\phi^-\overline{\phi}^0
-\phi^-\overline{\phi}^0$) belong to $I=0$ states. 
The constraint (6.10) leads the $\lambda_3$- the $\eta_3$-terms in 
the potential to 
$$
\frac{1}{2}\lambda_3\left[
{\rm Tr}(\overline{\phi}_{oct}\overline{\phi}_{oct})-
\overline{\phi}_s\overline{\phi}_s\right]
\cdot\left[{\rm Tr}(\phi_{oct}\phi_{oct})-\phi_s\phi_s\right]
\ , \eqno(6.14)
$$
where $\overline{\phi}\overline{\phi}\cdot\phi\phi$ denotes the $I=0$ 
component in $(I=1)\times (I=1)$, i.e., 
$$
\overline{\phi}\overline{\phi}\cdot\phi\phi=
\phi^-\phi^-\cdot\phi^+\phi^+ + 
\frac{\phi^-\overline{\phi}^0+\overline{\phi}^0\phi^-}
{\sqrt{2}}\cdot
\frac{\phi^+\phi^0+\phi^0\phi^+}{\sqrt{2}}+
\overline{\phi}^0\overline{\phi}^0\cdot\phi^0\phi^0
\ . \eqno(6.15)
$$
Then, the physical Higgs boson masses are given by 
$$
\begin{array}{l}
m_{H1}^2\equiv m^2(H_1^0)=2(\lambda_1+\lambda_2)v_0^2 \ , \\
m_{H2}^2\equiv m^2(H_2^0)=m^2(\chi_2^0)=2\lambda_3 v_0^2 \ , \\
m_{H3}^2\equiv m^2(H_3^0)=m^2(H_i^{0j})
=m^2(\chi_3^0)=m^2(\chi_i^{0j})=\overline{\xi} v_0^2 \ , \\
m^2(\chi_2^{\pm})=-(\lambda_2+\eta_2-\lambda_3)v_0^2 \ , \\
m^2(\chi_3^{\pm})=-\left( \lambda_2+\frac{1}{2}\eta_2-\overline{\xi}
\right) v_0^2 \ . 
\end{array} \eqno(6.16)
$$

The mass difference between $P_i^j$ and $\overline{P}_i^j$, 
$\Delta m_P=m(P_i^j)-m(\overline{P}_i^j)$, is given by [12]
$$
\Delta m_P =  \eta_{QCD} B_P f_P^2 m_P \left[
\left(\frac{m_P}{m_i+m_j}\right)^2 -\frac{1}{6}\right] 
\frac{1}{3}\left(\frac{1}{m_{H2}^2}-\frac{1}{m_{H3}^2}
\right)K_{ij} \ , \eqno(6.17)
$$
where  $\eta_{QCD}$ ia a QCD correction 
factor from hard gluon exchange, $B_P$ is a parameter that 
characterizes the inaccuracy of 
the vacuum insertion approximation, and 
$$
K_{ij}=
\frac{m_im_j}{v_0^2}\sum_k \left(
\frac{1}{z_k^2}+\frac{z_k-z_l-z_m}{z_1z_2z_3}\right) 
(U_i^kU_j^{k*})^2 \ . \eqno(6.18)
$$
Although the ($\overline{u}c$) and ($\overline{d}s$) currents involve 
the small factors $m_um_c/v_0^2\simeq2.8\times10^{-7}$ and 
$m_dm_s/v_0^2\simeq6.5\times10^{-8}$, 
respectively, the contributions of them to the $K_L$-$K_S$ and 
$D^0$-$\overline{D}^0$ mass differences are not negligible 
because $K_{ij}$ are given by $K_{12}\simeq |(U_L^d)_2^1|^2/z^2_1$ for 
$\Delta m_K$ and $K_{12}\simeq |(U_L^u)_2^1|^2/z_1^2$ for 
$\Delta m_D$, and the factor $1/z_1^2$ takes a large value 
$1/z_1^2=3.685\times10^3$. 
The experimental values [6] 
$m_{K_L}-m_{K_S}=(3.510\pm 0.018)\times 10^{-12}$ MeV, 
$|m_{D_1^0}-m_{D_2^0}|<20\times 10^{10}$ $\hbar$s$^{-1}$, 
$m_{B_H^0}-m_{B_L^0}=(0.51\pm 0.06)\times 10^{12}$ $\hbar$s$^{-1}$, 
and 
$m_{B_{sH}^0}-m_{B_{sL}^0}> 1.8\times 10^{12}$ $\hbar$s$^{-1}$ lead to 
the constraints
$$
\left(1/m_{H_2}^{2}-1/m_{H_3}^{2}\right)^{-1/2} \simeq \left(\eta_{QCD}^K
\right)^{1/2} |(U_L^d)_1^1 (U_L^d)_2^{1*}|\times 32 \ {\rm TeV}\ , 
\eqno(6.19)
$$ 
$$
\left(1/m_{H_2}^{2}-1/m_{H_3}^{2}\right)^{-1/2} > \left(\eta_{QCD}^D
\right)^{1/2} |(U_L^u)_1^1 (U_L^u)_2^{1*}| \times 16 \ {\rm TeV} 
\ , \eqno(6.20)
$$ 
$$
\left(1/m_{H_2}^{2}-1/m_{H_3}^{2}\right)^{-1/2} \simeq \left(\eta_{QCD}^B
\right)^{1/2} |(U_L^d)_1^1 (U_L^d)_3^{1*}|\times 38 \ {\rm TeV} 
\ , \eqno(6.21)
$$ 
and
$$
\left(1/m_{H_2}^{2}-1/m_{H_3}^{2}\right)^{-1/2} < \left(\eta_{QCD}^{Bs}
\right)^{1/2} |(U_L^d)_2^1 (U_L^d)_3^{1*}|\times 88 \ {\rm TeV}  
\ , \eqno(6.22)
$$ 
respectively, where we, for simplicity, have put the other contributions 
to $P^0$-$\overline{P}^0$ mixing zero, and,  we have used 
$B_K=0.65$ [13] and $f_K=0.160$ GeV [6] in (6.19) and 
$f_D B_D^{1/2}=f_B B_B^{1/2}=f_{Bs} B_{Bs}^{1/2}=0.2$ GeV in (6.20) -- 
(6.22).

The numerical estimates of the Higgs boson masses depend on 
the structures of $U_L^u$ and $U_L^d$.  From the constraint 
$V_{us}=(U_L^u U_L^{d\dagger})_1^2\simeq 0.22$, 
we cannot consider a mass matrix model which provides 
$(U_L^u)_1^2\simeq 0$ and $(U_L^d)_1^2\simeq 0$ simultaneously.
We suppose $|(U_L^u)_1^2|\sim |(U_L^d)_1^2|$.
If we take $|(U_L^d)_1^1(U_L^d)_1^2|\simeq 0.22$ by way of trial and 
$\eta_{QCD} \simeq 1$, the constraint (6.19) predicts 
$\left(1/m_{H_2}^{2}-1/m_{H_3}^{2}\right)^{-1/2} \simeq 7.1$ TeV.
Only when $m_{H2}\simeq m_{H3}$, the FCNC processes are 
highly suppressed.  
Since we have known $m_{H3}>2.4$ TeV from the data of 
$K_L\rightarrow e^\pm \mu^\mp$,
we cannot take too low a value of $m_{H2}$. 
For example, for $m_{H3}\simeq 2.5$ TeV, 
the mass difference must be a very small value 
$m_{H3}-m_{H2}\simeq 0.14$ TeV.

\vglue.3in

\noindent
{\bf 7. Production and decays of new Higgs bosons}

As we stated in Sect.4 and Sect.6, as far as the Higgs boson $H_1^0$ is 
concerned, its interactions with electroweak gauge bosons and with 
light fermions (quarks and leptons) are exactly the same ones as the 
physical Higgs boson $H_{SM}^0$ in the standard model.  From (4.16), 
the decays $H_1^0\rightarrow H_k^0H_k^0$, 
$\chi_k^0\chi_k^0$ $(k=2,3)$, $(H^0)_i^j(H^0)_j^i$ and 
$(\chi^0)_i^j(\chi^0)_j^i$ $(i\neq j)$ are forbidden, so that 
the dominant decay mode is  $H_1^0\rightarrow b\overline{b}$.
Therefore, in the present model, it is hard to distinguish 
the Higgs boson $H_1^0$ from $H_{SM}^0$ in the standard model. 

The most distinguishable ones from the physical Higgs bosons in 
the standard model and/or in the conventional multi-Higgs model 
are $(H^0)_i^j$ and $(\chi^0)_i^j$ $(i\neq j)$. 
If we suppose that they have masses of a several hundred GeV, 
we may expect a production 
$$
e^+ + e^- \rightarrow Z^* \rightarrow (H^0)_i^j \ \ \ \ \ \
+ \ (\chi^0)_j^i \ , \eqno(7.1)
$$
$$\hspace{2.8cm}\hookrightarrow f_i+\overline{f}_j \ \ \ \ \ \ 
\hookrightarrow f_j+\overline{f}_i
$$
in a super  $e^+e^-$ linear collider in the near future. 
Unfortunately, as discussed in Sect.5 and Sect.6, their masses must be
larger than a few TeV, so that we cannot expect the 
observation in $e^+ e^-$ collider.

Only a chance of the observation of our Higgs bosons $\phi_i^j$ is 
in a production 
$$ u + q (\overline{q}) \rightarrow t + (\phi)_1^3 +q (\overline{q}) 
\ \ \ \ \ (q=u, d, s) \ ,\eqno(7.2) $$
at a super hadron collider with several TeV beam energy,
for example, at LHC, 
because the coupling  $a_{tu}$ ($b_{tu}$) is sufficiently large to 
produce (7.2): 
$$
a_{tu} \simeq \frac{m_t}{v_3} +\frac{m_u}{v_1}=1.029+ 0.002
. \eqno(7.3)
$$

The dominant decay modes of $(H^0)_2^{3}$ and $(H^0)_1^{3}$ 
are hadronic ones, i.e., $(H^0)_2^{3}\rightarrow c\overline{t}$, 
$s\overline{b}$ and $(H^0)_1^{3}\rightarrow u\overline{t}$, 
$d\overline{b}$ . 
Only for $(H^0)_1^{2}$, which is produced by the reaction 
$u+q\rightarrow c+(H^0)_1^2 +q$, 
the leptonic mode $(H^0)_1^{2}\rightarrow e^-\mu^+$ 
has a visible branching ratio: 
$$
\Gamma(H_1^{2}\rightarrow u\overline{c}):
\Gamma(H_1^{2}\rightarrow d\overline{s}) 
:\Gamma(H_1^{2}\rightarrow e^-\mu^+)
$$
$$
\simeq 3\left[\left(\frac{m_c}{v_2}\right)^2+
\left(\frac{m_u}{v_1}\right)^2\right] : 
3\left[\left(\frac{m_s}{v_2}\right)^2
+\left(\frac{m_d}{v_1}\right)^2\right] : 
\left[\left(\frac{m_{\mu}}{v_2}\right)^2
+\left(\frac{m_e}{v_1}\right)^2\right]
$$
$$
= 73.5\% : 24.9\% : 1.6\% ,
\eqno(7.4)
$$
where we have used an approximate relation $U_L^f\simeq {\bf 1}$
and have taken the quark mass values $m_q(\mu)$ at 
$\mu=1$ GeV, $m_u=5.6$ MeV, $m_d=9.9$ MeV, $m_s=199$ MeV and 
$m_c=1.49$ GeV, as the quark masses inside ordinary hadrons.

\vglue.3in

\noindent
{\bf 8. Conclusion}

In conclusion, inspired by the phenomenological success of the 
charged lepton mass relation (1.4), we have proposed a model 
with U(3)$_{family}$ nonet Higgs bosons $\phi_L$ and $\phi_R$ and 
vector-like heavy leptons $F$ ($F=U,D,N,E$) correspondingly 
to ordinary quarks and leptons $f$ ($f=u,d,\nu,e$), 
and have investigated its possible new physics. 

The charged lepton mass relation (1.4) can derive only when 
the potential $V(\phi)$ takes a special form (3.1), 
which satisfies ``U(3)-family nonet" ansatz. 
In order to avoid massless physical Higgs bosons, 
we must consider a term which explicitly breaks U(3)-family 
symmetry, (3.13) [or (3.16)].

In the low energy phenomenology, only the light Higgs boson 
$\phi_L$ plays a role.
Of the 36 components of our Higgs boson $\phi_L$, the three, 
$\chi_1^\pm$ and $\chi_1^0$, are eaten by the gauge bosons 
$W^\pm$ and $Z^0$, respectively.
The neutral Higgs boson $H_1^0$ has the same interactions 
with fermions and electroweak gauge bosons, so that it is 
hard to distinguish our Higgs boson $H_1^0$ from the neutral 
Higgs boson in the standard model experimentally.

If we take the ${\xi}\rightarrow\infty$ limit in the 
explicit U(3)-family symmetry breaking term $V_{SB}$,
the Higgs bosons which have finite masses become only 
$H_1^0$, $H_2^0$, $\chi_2^0$ and $\chi_2^\pm$, so that 
the model becomes similar to the two-Higgs-doublet model. 
However, differently from the conventional 
two-Higgs-doublet model, our Higgs bosons 
$H_2^0$ and $\chi_2^0$ 
can contribute to the flavor-changing neutral current
processes, so that their masses must be larger than $10^2$ TeV.
We think that such a case is unnatural.

For the case of a finite $\overline{\xi}$, we have 33 physical Higgs 
bosons given by (4.5), (4.11) and (4.12).
In order to suppress the rare decay modes  
$K_L \rightarrow \mu^\pm e^\mp$, $K^+ \rightarrow \pi^+e^-\mu^+$, 
and so on, we must put the constraint (5.20), which leads to 
$m(H_3)\simeq m(\chi_3)>2.4$ TeV for $m_H\simeq m_\chi$. 
In order to suppress the FCNC, the masses of $H_2^0$ and $\chi_2^0$ 
must be, in general, larger than $10^2$ TeV. 
Only the case which gives an acceptably lower values of the Higgs boson
masses is the case $m(H)=m(\chi)$ and $m_{H2}\simeq m_{H3}$. 
Then, we can expect our Higgs bosons with masses of 
2.5 TeV (except for $H_1^0$). 
In a top pair production at LHC, we may expect 
an observation of $t\overline{t}$ pair with a large $p_T$, 
due to the production $u+q \rightarrow t+\phi_1^3 + q$ and 
the subsequent decay $\phi_1^3 \rightarrow u+\overline{t}$.

\vglue.3in

\centerline{\bf Acknowledgments}

The problem of the flavor-changing neutral currents in the present model
was pointed out by Professor K.~Hikasa. 
The authors would sincerely like to thank  him 
for valuable comments.
This work was supported by the Grant-in-Aid for Scientific Research, 
Ministry of Education, Science and Culture, Japan (No.06640407).


\vglue.3in

\begin{center}
{\bf Appendix A}
\end{center}

More general potential form of $V_{nonet}$ is given by 
$$
V_{nonet}=[{\rm r.h.s.\ of\ (3.2)}] 
+\sum_{i,j,k,l}\left[\frac{1}{2}\lambda_4 
(\overline{\phi}_i^j \phi_j^k)(\overline{\phi}_k^l \phi_l^i) \right.
$$
$$\left.
+\frac{1}{2}\lambda_5 
(\overline{\phi}_i^j \phi_l^i)(\overline{\phi}_k^l \phi_j^k) 
+\frac{1}{2}\lambda_6 
(\overline{\phi}_i^j \phi_j^k)(\overline{\phi}_l^i \phi_k^l) 
+\frac{1}{2}\lambda_7 
(\overline{\phi}_i^j \phi_k^l)(\overline{\phi}_j^k \phi_l^i)\right] \ . 
\eqno(A.1)
$$ 
Then, for $\mu^2<0$, 
conditions for minimizing the potential (3.1) are as follows:
$$
\hspace*{-14.5cm}\lefteqn{
\left[\mu^2+(\lambda_1+\lambda_2){\rm Tr}(v^{\dagger}v)\right]v_s^*
+ \lambda_3{\rm Tr}(v^{\dagger}v^{\dagger})v_s
+ (\lambda_4+\lambda_5+\lambda_6+\lambda_7)
\frac{1}{\sqrt{3}}{\rm Tr}(v^{\dagger}v^{\dagger}v)
} $$
$$
+ (\eta_1+\eta_2){\rm Tr}(v^{\dagger}_{oct}v_{oct})v_s^*
+ 2\eta_3^*{\rm Tr}(v^{\dagger}_{oct}v^{\dagger}_{oct})v_s=0 \ , \eqno(A.2)
$$
$$
\hspace*{-14.5cm}\lefteqn{
\left[\mu^2+(\lambda_1+\lambda_2){\rm Tr}(v^{\dagger}v)\right]
v_{oct}
+ \lambda_3{\rm Tr}(v^{\dagger}v^{\dagger})v_{oct}
+ (\lambda_4+\lambda_5)(v^{\dagger}vv^{\dagger})_{oct}
}$$
$$
+ \frac{1}{2}(\lambda_6+\lambda_7)(v^{\dagger}v^{\dagger}v+vv^{\dagger}
v^{\dagger})_{oct}
+ (\eta_1+\eta_2)v_s^*v_s^*v^{\dagger}_{oct}
+ 2\eta_3v_s^*v_s^*v_{oct}=0 \ , \eqno(A.3)
$$
and the similar equations with $v\leftrightarrow v^\dagger$ 
($v_{oct}\leftrightarrow v^\dagger_{oct}$ and 
$v_s\leftrightarrow v_s^\dagger$), 
where $A_{oct}$ means an octet part of the $3\times 3$ matrix $A$, i.e., 
$A_{oct}=A-(1/3){\rm Tr}(A)\, {\bf 1}$.
Only when
$$
\lambda_4+\lambda_5+\lambda_6+\lambda_7=0 \ , \eqno(A.4)
$$
we can obtain the desirable relation (3.6).

What is of great interest to us is whether these terms ($\lambda_4$ - 
$\lambda_7$ terms) can generate additional masses of the neutral Higgs 
bosons $H_i^{0j}$ ($i\neq j$) or not, because these bosons become to 
massless in the limit of $\overline{\xi}\rightarrow 0$.
Unfortunately, these terms cannot contribute to the masses except for 
those of $\chi_i^{\pm j}$ and $\chi_i^{0 j}$ ($i\neq j$), 
because the $\lambda_4$ - $\lambda_7$ terms still respect the 
SU(3) family symmetry.
Therefore, the exitstnce of Goldstone bosons cannot be avoided by 
the introduction of these terms. 
The existence of the $\lambda_4$ - $\lambda_7$ terms does not improve 
the situation of our model and only makes our study intricate, 
so that we have omitted the study of the $\lambda_4$ - $\lambda_7$ terms 
from the present studies.

\vglue.3in

\begin{center}
{\bf Appendix B}
\end{center}

Since the parameters $z_i=v_i/v_0$ satisfy the relations
$$
z_1^2 + z_2^2 + z_3^2 = 1 \ , \eqno(B.1)
$$
$$
z_1 + z_2 + z_3 = \sqrt{\frac{3}{2}} \ , \eqno(B.2)
$$
we find the relations
$$
z_1 z_2 + z_2 z_3 + z_3 z_1 = \frac{1}{4} \ , \eqno(B.3)
$$
$$
z_iz_j = \frac{1}{4}-\sqrt{\frac{3}{2}}z_k + z_k^2 \ , \eqno(B.4)
$$
where $(i,j,k)$ are cyclic indices of $(1,2,3)$.

In the present seesaw-type mass matrix model, the values $v_i^2$ are 
proportional to the charged lepton masses 
$m_i^e = (m_e, m_{\mu}, m_{\tau})$, the values $z_i$ are 
given by 
$$
z_i=[m_i^e/(m_e+m_{\mu}+m_{\tau})]^{1/2} \ , \eqno(B.5)
$$
i.e.,
$$
\begin{array}{ccc}
\left( 
\begin{array}{c}
z_1 \\
z_2 \\
z_3
\end{array} \right)
& = & 
\left( 
\begin{array}{l}
0.016473 \\
0.23687 \\
0.97140
\end{array} \right)
\end{array}
\ , \eqno(B.6)
$$
so that
$$
\begin{array}{ccc}
\left( 
\begin{array}{c}
x_1 \\
x_2 \\
x_3
\end{array} \right)
& = & 
\left( 
\begin{array}{l}
-0.55405 \\
-0.24237 \\
+0.79642
\end{array} \right)
\end{array}
$$
$$
=\cos(\frac{\pi}{4}+\delta)\frac{1}{\sqrt{6}}\left(
\begin{array}{c}
-2 \\
1 \\
1
\end{array}\right)
+\sin(\frac{\pi}{4}+\delta)\frac{1}{\sqrt{2}}\left(
\begin{array}{c}
0 \\
-1 \\
1
\end{array}\right) \ , \ \ \ (\delta=2.268^\circ)
\ . \eqno(B.7)
$$
The expression (B.7) suggests that the Higgs boson state $\phi_x$ 
is almost given by a 
$45^\circ$-mixing between $\lambda_3$- and $\lambda_8$-components 
of SU(3)$_{family}$.
At the present stage, the parameter $\delta$ is pure phenomenological one. 
If we can give only the value of $\delta$, we can fix the values of 
$(x_1,x_2,x_3; y_1,y_2,y_3; z_1,z_2,z_3)$. 
The open question why $\delta$ takes such a value will be answered in a 
future theory.


\vglue.3in

\vglue.3in
\newcounter{0000}
\centerline{\bf References and Footnotes}
\begin{list}
{[~\arabic{0000}~]}{\usecounter{0000}
\labelwidth=0.8cm\labelsep=.1cm\setlength{\leftmargin=0.7cm}
{\rightmargin=.2cm}}
\item The ``family symmetry" is also called a ``horizontal symmetry":
K.~Akama and H.~Terazawa, Univ.~of Tokyo, report No.~257 (1976) 
(unpublished); T.~Maehara and T.~Yanagida, Prog.~Theor.~Phys. {\bf 60}, 
822 (1978); F.~Wilczek and A.~Zee, Phys.~Rev.~Lett. {\bf 42}, 421 (1979); 
A.~Davidson, M.~Koca and K.~C.~Wali, Phys.~Rev. {\bf D20}, 1195 (1979); 
J.~Chakrabarti, Phys.~Rev. {\bf D20}, 2411 (1979). 
\item M.~Kobayashi and T.~Maskawa, Prog.~Theor.~Phys. {\bf 49}, 652 (1973).
\item The seesaw mechanism has originally proposed for the purpose of 
explaining why neutrino masses are so invisibly small:
M.~Gell-Mann, P.~Rammond and R.~Slansky, in {\it Supergravity}, 
edited by P.~van Nieuwenhuizen and D.~Z.~Freedman (North-Holland, 
1979); T.~Yanagida, in {\it Proc. Workshop of the Unified Theory and 
Baryon Number in the Universe}, edited by A.~Sawada and A.~Sugamoto 
(KEK, 1979); R.~Mohapatra and G.~Senjanovic, Phys.~Rev.~Lett. 
{\bf 44}, 912 (1980). 
For  applications of the seesaw mechanism 
to the quark mass matrix, see, for example, 
Z.~G.~Berezhiani, Phys.~Lett. {\bf 129B}, 99 (1983);
Phys.~Lett. {\bf 150B}, 177 (1985);
D.~Chang and R.~N.~Mohapatra, Phys.~Rev.~Lett. {\bf 58}, 1600 
(1987); 
A.~Davidson and K.~C.~Wali, Phys.~Rev.~Lett. {\bf 59}, 393 (1987);
S.~Rajpoot, Mod.~Phys.~Lett. {\bf A2}, 307 (1987); 
Phys.~Lett. {\bf 191B}, 122 (1987); Phys.~Rev. {\bf D36}, 1479 (1987);
K.~B.~Babu and R.~N.~Mohapatra, Phys.~Rev.~Lett. {\bf 62}, 1079 (1989); 
Phys.~Rev. {\bf D41}, 1286 (1990); 
S.~Ranfone, Phys.~Rev. {\bf D42}, 3819 (1990); 
A.~Davidson, S.~Ranfone and K.~C.~Wali, Phys.~Rev. 
{\bf D41}, 208 (1990); 
I.~Sogami and T.~Shinohara, Prog.~Theor.~Phys. {\bf 66}, 1031 (1991);
Phys.~Rev. {\bf D47}, 2905 (1993); 
Z.~G.~Berezhiani and R.~Rattazzi, Phys.~Lett. {\bf B279}, 124 (1992);
P.~Cho, Phys.~Rev. {\bf D48}, 5331 (1994); 
A.~Davidson, L.~Michel, M.~L,~Sage and  K.~C.~Wali, Phys.~Rev. 
{\bf D49}, 1378 (1994); 
W.~A.~Ponce, A.~Zepeda and R.~G.~Lozano, Phys.~Rev. {\bf D49}, 4954 
(1994).
\item Y.~Koide, Mod.~Phys.~Lett. {\bf A5}, 2319 (1990).
\item Y.~Koide, Lett.~Nuovo Cimento {\bf 34}, 201 (1982); Phys.~Lett. 
{\bf B120}, 161 (1983);  Phys.~Rev. {\bf D28}, 252 (1983).
\item Particle data group, Phys.~Rev. {\bf D50}, 1173 (1994).
\item Y.~Koide, a talk presented at the INS Workshop ^^ ^^ Physics of 
$e^{\dagger}e^{-}$, 
$e^{-}\gamma$ and $\gamma\gamma$ collisions at linear accelerators", INS, 
University of Tokyo, December 20--22, 1994, to be published in Proceedings 
edited by  Z.~Hioki,  T.~Ishii and R.~Najima.
The prototype of this model was investigated by Fusaoka and one of 
the authors (Y.K.): 
Y.~Koide and H.~Fusaoka, US-94-02, 1994 (hep-ph/9403354), (unpublished).
However, their Higgs potential leads to massless 
physical Higgs bosons, so that it brings some troubles into the theory.
In the present paper, the global symmetry U(3)$_{family}$ will be 
broken explicitly, and not spontaneously, so that massless physical 
Higgs bosons will not appear.
\item H.~Harari, H.~Haut and J.~Weyers, Phys.~Lett. {\bf B78},
459 (1978).
%
%
\item Y.~Koide and H.~Fusaoka, Preprint US-95-03 and AMU-95-04 (1995) 
(hep-ph 9505201), to be published in Z.~Phys. C.
\item H.~Harari, H.~Haut and J.~Weyers, Phys.~Lett. {\bf B78}
(1978) 459;
T.~Goldman, in {\it Gauge Theories, Massive Neutrinos and 
Proton Decays}, edited by A.~Perlumutter (Plenum Press, New York, 
1981), p.111;
T.~Goldman and G.~J.~Stephenson,~Jr., Phys.~Rev. {\bf D24} 
(1981) 236; 
Y.~Koide, Phys.~Rev.~Lett. {\bf 47} (1981) 1241; Phys.~Rev. 
{\bf D28} (1983) 252; {\bf 39} (1989) 1391;
C.~Jarlskog, in {\it Proceedings of the International Symposium on 
Production and Decays of Heavy Hadrons}, Heidelberg, Germany, 1986
edited by K.~R.~Schubert and R. Waldi (DESY, Hamburg), 1986, p.331;
P.~Kaus, S.~Meshkov, Mod.~Phys.~Lett. {\bf A3} (1988) 1251; 
Phys.~Rev. {\bf D42} (1990) 1863;
L.~Lavoura, Phys.~Lett. {\bf B228} (1989) 245; 
M.~Tanimoto, Phys.~Rev. {\bf D41} (1990) 1586;
H.~Fritzsch and J.~Plankl, Phys.~Lett. {\bf B237} (1990) 451; 
Y.~Nambu, in {\it Proceedings of the International Workshop on 
Electroweak Symmetry Breaking}, Hiroshima, Japan, (World 
Scientific, Singapore, 1992), p.1.
\item V.~Barger, J.~L.~Hewett and R.~J.~N.Phillips, Phys.~Rev. {\bf D41},
3421 (1990); L.~Hall and S.~Weinberg, Phys.~Rev. {\bf D48}, R979 (1993);
Y.~L.~Wu and L.~Wolfernstein, Phys.~Rev.~Lett. {\bf 73}, 1762 (1994).
\item Note that the effective interaction (6.11) has not the form 
$[(\overline{f}_if_j)^2+(\overline{f}_i\gamma_5 f_j)^2]$ 
in the conventional model, 
but the form $[(\overline{f}_if_j)^2-(\overline{f}_i\gamma_5 f_j)^2]$, 
so that the factor 5/6 in the conventional model is replaced with 
the factor $[(\cdots)^2-1/6]$ as given in (6.17). 
See, for example, B.~McWilliams and O.~Shanker, Phys.~Rev. {\bf D22}, 
2853 (1980).
\item A.~J.~Buras, M.~Jamin and P.~H.~Weisz, Nucl.~Phys. {\bf B347}, 
491 (1990).
\end{list}


\newpage

Table I.  Quantum numbers of fermions and Higgs bosons

\vglue.1in
\begin{center}
\begin{tabular}[t]{|c|c|c|c|c|} \hline 
      &  $Y$ & SU(2)$_L$    & SU(2)$_R$  & U(3)$_{family}$  \\
 \hline  \hline 
 $f_L$  & $(\nu, \: e)_L^{Y=-1}$, $(u, \: d)_L^{Y=1/3}$ 
& {\bf 2} & {\bf 1} & {\bf 3}  \\
 $f_R$ & $(\nu, \: e)_R^{Y=-1}$, $(u, \: d)_R^{Y=1/3}$   
& {\bf 1} & {\bf 2} & {\bf 3}   \\
$F_L$ & $N_L^{Y=0}$, $E_L^{Y=-2}$, $U_L^{Y=4/3}$, $D_L^{Y=-2/3}$ 
& {\bf 1} & {\bf 1} & {\bf 3}   \\   
$F_R$ & $N_R^{Y=0}$, $E_R^{Y=-2}$, $U_R^{Y=4/3}$, $D_R^{Y=-2/3}$ 
& {\bf 1} & {\bf 1} & {\bf 3} \\  \hline 
$\phi_L$ & $(\phi^+,\phi^0)_L^{Y=1}$ & {\bf 2} & {\bf 1} & {\bf 8}+{\bf 1} \\ 
$\phi_R$ & $(\phi^+,\phi^0)_R^{Y=1}$ &{\bf 1} & {\bf 2} & {\bf 8}+{\bf 1} \\
$\Phi_F$ & $\Phi_0^{Y=0}$, $\Phi_X^{Y=0}$ & {\bf 1} & {\bf 1} & {\bf 1},
\ {\bf 8}+{\bf 1}  \\ \hline
\end{tabular} 
\end{center}
\vspace{2cm}
\setlength{\unitlength}{0.7mm}
\begin{picture}(160,100)(0,-25)
\linethickness{0.5mm}
\thicklines\put(0,30){\line(1,0){160}}
\thicklines\multiput(20,30)(40,0){4}{\vector(1,0){0}}
\multiput(40,30)(0,2){10}{\line(0,1){1}}
\multiput(80,30)(0,2){10}{\line(0,1){1}}
\multiput(120,30)(0,2){10}{\line(0,1){1}}
\multiput(40,30)(40,0){3}{\circle*{2}}
\put(0,20){\makebox(40,10)[b]{$f_L$}}
\put(40,20){\makebox(40,10)[b]{$F_R$}}
\put(80,20){\makebox(40,10)[b]{$F_L$}}
\put(120,20){\makebox(40,10)[b]{$f_R$}}
\put(30,20){\makebox(20,10)[c]{$y_f^L$}}
\put(70,20){\makebox(20,10)[c]{$y_F$}}
\put(110,20){\makebox(20,10)[c]{$y_f^R$}}
\put(0,10){\makebox(40,10)[b]{({\bf 2, 1, 3})}}
\put(40,10){\makebox(40,10)[b]{({\bf 1, 1, 3})}}
\put(80,10){\makebox(40,10)[b]{({\bf 1, 1, 3})}}
\put(120,10){\makebox(40,10)[b]{({\bf 1, 2, 3})}}

\put(15,55){\makebox(40,10)[t]{$\langle\phi_L^0\rangle_0\equiv m_L/y_f^L$}}
\put(60,55){\makebox(40,10)[t]{$\langle\Phi_F\rangle_0\equiv M_F/y_F$}}
\put(105,55){\makebox(40,10)[t]{$\langle\phi_R^0\rangle_0\equiv m_R/y_f^R$}}
\put(20,65){\makebox(40,10)[t]{({\bf 2, 1, 8+1})}}
\put(60,65){\makebox(40,10)[t]{({\bf 1, 1, 1})}}
\put(100,65){\makebox(40,10)[t]{({\bf 1, 2, 8+1})}}
\multiput(30,40)(40,0){3}{\makebox(20,20){{\bf $\times$}}}
\end{picture}

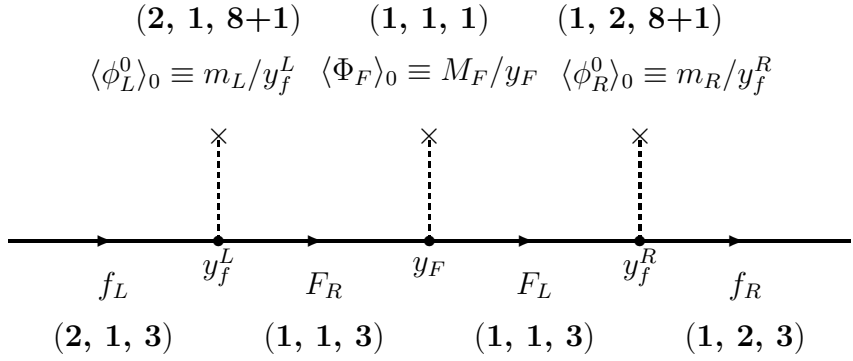
Fig. 1.  Mass generation of quarks and leptons $f$


\end{document}